# Chapter 4

# Electromagnetic Wave Source Conditions[1]

*Ardavan Oskooi and Steven G. Johnson*

## 4.1 OVERVIEW

This chapter discusses the relationships between current sources and the resulting electromagnetic waves in FDTD simulations. First, the "total-field/scattered-field" approach to creating incident plane waves is reviewed and seen to be a special case of the well-known *principle of equivalence* in electromagnetism: this can be used to construct "equivalent" current sources for any desired incident field, including waveguide modes. The effects of dispersion and discretization are discussed, and a simple technique to separate incident and scattered fields is described in order to compensate for imperfect equivalent currents. The important concept of the *local density of states* (LDOS) is reviewed, which elucidates the relationship between current sources and the resulting fields, including enhancement of the LDOS via mode cutoffs (Van Hove singularities) and resonant cavities (Purcell enhancement). We also address various other source techniques such as covering a wide range of frequencies and incident angles in a small number of simulations for waves incident on a periodic surface, sources to excite eigenmodes in rectangular supercells of periodic systems, moving sources, and thermal sources via a Monte Carlo/Langevin approach.

## 4.2 INCIDENT FIELDS AND EQUIVALENT CURRENTS

A common problem in FDTD simulations is to compute the interaction of a given incident wave with some geometry and materials in a localized region. For example, the incident wave could be a plane wave in vacuum (e.g., to compute the scattered and absorbed field for some isolated object), or the incident wave could be a propagating mode of a waveguide (e.g., to compute the transmission or reflection around a waveguide bend or through some other device). One cannot simply specify the fields near the boundary of a simulation — this would be a "hard" source that nonphysically scatters waves (e.g., reflected waves) that impinge on the source [1]. Instead, it is desirable to specify *equivalent electric and magnetic currents* (also called "transparent sources" [1]) that produce the desired incident wave, but which are transparent to other waves by the linearity of the Maxwell equations (at least in a linear incident medium).

---

[1] In this chapter the symbol $i$ is used to designate $\sqrt{-1}$, rather than the symbol $j$; and a phasor is denoted as $e^{-i\omega t}$.

In fact, there is a simple general prescription for deriving such currents from *any* incident field based on one of the fundamental theorems of electromagnetism: the *principle of equivalence* [2–4], a precise formalization of *Huygens' principle* that fields on a wavefront can be treated as sources. In the context of FDTD, these equivalent currents enable the "total-field/scattered-field" approach [1], but many other applications are possible.

### 4.2.1  The Principle of Equivalence

The derivation in this section is essentially equivalent to the usual derivation of the total-field/scattered-field approach, but instead of first discretizing the equations and then writing out the FDTD component equations individually, we start with the analytical Maxwell's equations and employ a compact notation that allows us to look at all the equations together. Not only does this shorten the derivation, but it also makes clear the applicability of the approach to any linear incident medium (such as a waveguide, including inhomogeneous and bianisotropic media [4]), and highlights the explicit identification of the source terms as the electric and magnetic currents of the principle of equivalence.

It is convenient to write Maxwell's equations compactly in terms of the six-component electric (***E***) and magnetic (***H***) field vector, $\psi$, and the six-component electric (***J***) and magnetic (***K***) current vector, $\xi$, in which case the equations in linear media can be written as:

$$\underbrace{\begin{bmatrix} & \nabla \times \\ -\nabla \times & \end{bmatrix}}_{\mathbf{M}} \underbrace{\begin{bmatrix} E \\ H \end{bmatrix}}_{\psi} = \frac{\partial}{\partial t}(\psi + \chi * \psi) + \underbrace{\begin{bmatrix} J \\ K \end{bmatrix}}_{\xi} \qquad (4.1)$$

where we have chosen natural units in which $\varepsilon_0 = 1$ and $\mu_0 = 1$, and $\chi *$ denotes convolution with the $6 \times 6$ linear susceptibility tensor:

$$\chi = \begin{bmatrix} \varepsilon - 1 & \\ & \mu - 1 \end{bmatrix} \qquad (4.2)$$

in ordinary dielectric/magnetic media. Suppose we have a desired incident wave, $\psi_+$, which solves the source-free Maxwell's equations in an infinite "incident" medium, $\chi_+$ (e.g., vacuum or an infinite waveguide) with no scatterers: $\mathbf{M}\psi_+ = \partial(\psi_+ + \chi_+ * \psi_+)/\partial t$. Now, the question is to come up with *equivalent currents,* $\xi$, that generate the same incident field, $\psi_+$, in a *finite* domain, $\Omega$, within our computational space, into which we can then introduce scatterers or other inhomogeneities for interaction with the incident wave.

The derivation and application of the principle of equivalence can be thought of as a three-step process, illustrated in Fig. 4.1. First (a), we consider the incident wave, $\psi_+$, in an infinite medium, $\chi_+$, with space arbitrarily divided into an interior domain, $\Omega$, and an exterior, separated by an imaginary surface, $\partial\Omega$. Second (b), keeping the *same* infinite medium, $\chi_+$, we set $\psi = 0$ in the *exterior* and keep $\psi = \psi_+$ in the *interior* $\Omega$, and solve for the currents, $\xi$, which allow this *discontinuous* field, $\psi$, to solve Maxwell's equations: these are the *equivalent surface currents*. Third (c), given these currents, we can then truncate space [e.g., with perfectly matched layer (PML) absorbers], and introduce inhomogeneities to $\chi$ inside $\Omega$ in order to model the interaction of $\psi_+$ with these inhomogeneities. One can even include nonlinearities in the interior of $\Omega$ in step (c); the only real requirement is that the *incident* medium, $\chi_+$, be linear.

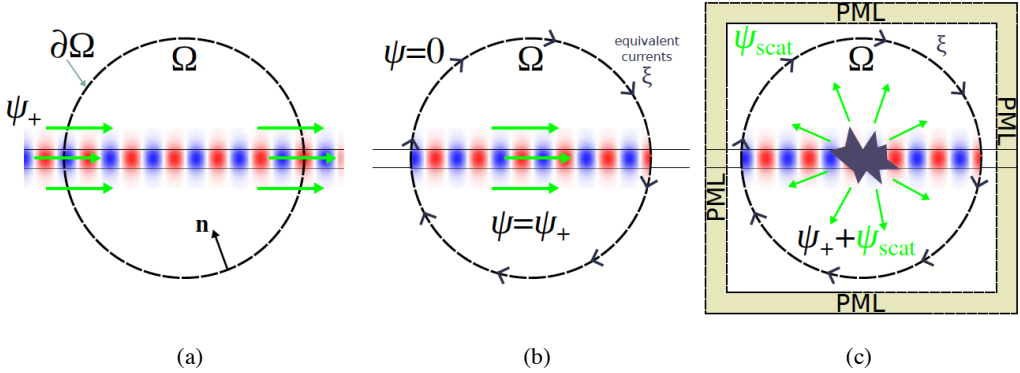

**Fig. 4.1** Schematic construction of equivalent electric and magnetic current sources, $\boldsymbol{\xi}$, to produce a desired incident field, $\boldsymbol{\psi}_+$, which denotes a 6-component composite electric and magnetic field vector. (a) The field, $\boldsymbol{\psi}_+$, in an infinite medium — here, a waveguide mode in an infinite waveguide — with space arbitrarily divided by an imaginary surface $\partial\Omega$ into an "interior" region, $\Omega$, and an "exterior." (b) We set the field to be $\boldsymbol{\psi} = 0$ in the exterior and $\boldsymbol{\psi} = \boldsymbol{\psi}_+$ in $\Omega$, and construct surface currents, $\boldsymbol{\xi}$, that allow this discontinuous $\boldsymbol{\psi}$ to satisfy Maxwell's equations. This is the *principle of equivalence*. (c) Given this $\boldsymbol{\xi}$, we can truncate the computation space with a PML absorber and insert a scatterer or some other object into $\Omega$. The current, $\boldsymbol{\xi}$, produces the desired incident field, $\boldsymbol{\psi}_+$, so that the field $\boldsymbol{\psi}$ inside $\Omega$ is the sum of $\boldsymbol{\psi}_+$ and a scattered field, while the field outside $\Omega$ is *only* the scattered field. This is the basis of the "total-field/scattered-field" approach in FDTD.

The key is the second step, depicted in Fig. 4.1(b). If we let:

$$\boldsymbol{\psi} = \begin{cases} \boldsymbol{\psi}_+ & \text{inside } \Omega \\ 0 & \text{outside } \Omega \end{cases} \quad (4.3)$$

then $\boldsymbol{\psi}$ clearly solves the source-free Maxwell's equations, $\mathbf{M}\boldsymbol{\psi} = \partial(\boldsymbol{\psi} + \boldsymbol{\chi}_+ * \boldsymbol{\psi})/\partial t$, in *both* the interior and exterior regions. Here, the only question is what happens at the surface, $\partial\Omega$. At this surface, the discontinuity of $\boldsymbol{\psi}$ has only one effect in Maxwell's equations: it produces a surface Dirac delta function, $\delta(\partial\Omega)$, in the spatial derivative, $\mathbf{M}\boldsymbol{\psi}$.[2] Hence, in order to satisfy Maxwell's equations with this $\boldsymbol{\psi}$, we must introduce a matching delta function on the right-hand side: a *surface current*, $\boldsymbol{\xi}$. In particular, if $\boldsymbol{n}$ is the unit inward-normal vector,[3] only the normal derivative, $\boldsymbol{n} \cdot \nabla$, contains a delta function (whose amplitude is the magnitude of the discontinuity). This implies a surface current:

$$\boldsymbol{\xi} = \begin{bmatrix} \boldsymbol{J} \\ \boldsymbol{K} \end{bmatrix} = \delta(\partial\Omega) \begin{bmatrix} \boldsymbol{n} \times \\ -\boldsymbol{n} \times \end{bmatrix} \boldsymbol{\psi}_+ = \delta(\partial\Omega) \begin{bmatrix} \boldsymbol{n} \times \boldsymbol{H}_+ \\ -\boldsymbol{n} \times \boldsymbol{E}_+ \end{bmatrix} \quad (4.4)$$

---

[2]That is, $\delta(\partial\Omega)$ is the distribution such that $\iiint \phi(\boldsymbol{x}) \delta(\partial\Omega) = \iint_{\partial\Omega} \phi(\boldsymbol{x})$ for any continuous test function, $\phi$.

[3]For simplicity, we assume a differentiable surface, $\partial\Omega$, so that its normal, $\boldsymbol{n}$, is well defined; but a surface with corners (e.g., a cubical domain) follows as a limiting case.

That is, there is a *surface electric current* given by the surface-tangential components, $n \times H_+$, of the incident magnetic field, and a *surface magnetic current* given by the components, $-n \times E_+$, of the incident electric field. These are the *equivalent currents* of the principle of equivalence (which can also be derived in other ways, traditionally from a Green's function approach [3, 4]). This principle has a long history [5–8] and far-reaching consequences in electromagnetism. For example, not only is it useful in constructing wave sources in FDTD, but it is also central to *integral-equation* formulations of electromagnetic scattering problems and the resulting boundary-element method (BEM) [9, 10].

Essentially the same derivation applied to the discretized curl operator, **M**, yields the source terms in the total-field/scattered field approach [1], with the source currents being a discrete or "Kronecker" delta function. Here, the connection to the equivalence principle usually goes unmentioned, despite the fact that this approach was originally derived in analogy with Huygens' principle [11]. Although this technique is most commonly applied to plane waves in free space, one can just as easily apply the derivation to produce a waveguide-mode source, as shown in Fig. 4.2. In this case, the incident fields are those of a waveguide mode. Since these are *exponentially localized* to the vicinity of the waveguide, we can in practice restrict the current sources, $\xi$, to a *short line segment crossing the waveguide*. An eigenmode current source excites only the desired rightward-propagating guided mode, whereas a simple constant-amplitude source produces a mixture of leftward- and rightward-propagating modes and radiated fields.

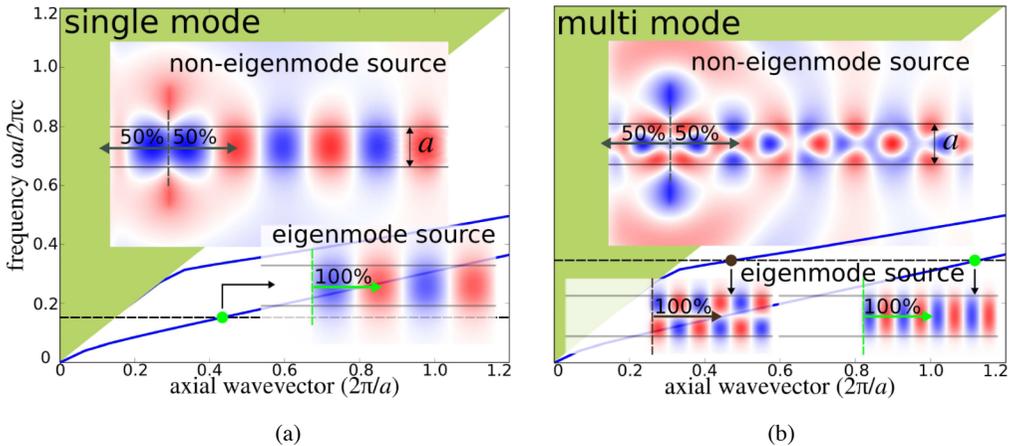

(a)                                          (b)

**Fig. 4.2** Comparison of different current sources in a two-dimensional (2-D) dielectric waveguide ($\varepsilon/\varepsilon_0 = 12$, width $a$) with out-of-plane **E** field. The dispersion relation, $\omega(k)$, is shown in the background of each panel, and the insets visualize the FDTD-computed $E_z$. (a) Sources at a frequency where the waveguide is single-mode, indicated by a dot on the $\omega(k)$ curve. Top inset: a simple out-of-plane electric-current source, **J**, of constant amplitude along a transverse cross-section of the waveguide (vertical dashed line), which excites both the waveguide mode and radiating fields in both directions. Bottom inset: equivalent-current source, $\xi$, from (4.4), having a transverse distribution consistent with the eigenmode computed in a mode solver, which excites only the right-going waveguide mode. (b) Sources at a frequency where the waveguide has two modes, indicated by two dots on the $\omega(k)$ curve. Top inset: a simple **J** source of constant amplitude along a transverse cross-section of the waveguide, which excites a superposition of both waveguide modes in addition to the radiating field. Bottom insets: equivalent-current source, $\xi$, having a transverse distribution consistent with either the fundamental or the 2nd-order eigenmode; in each case, we only excite the desired right-going mode.

### 4.2.2 Discretization and Dispersion of Equivalent Currents

There are two practical obstacles to exact implementation of equivalent-current sources in FDTD simulations: *discretization* and *dispersion*. If these are not accounted for precisely, then the currents, $\xi$, produce slight errors in $\psi_+$ in the interior, $\Omega$, and slightly nonzero fields in the exterior of $\Omega$. In practice, these slight errors can often be dealt with easily by postprocessing, as described in Section 4.3, but it is still important to understand how they arise and how (at least in principle) the errors could be eliminated.

The first potential source of error is discretization. As already mentioned, for FDTD we should, in principle, apply the equivalent-currents derivation to the discretized Maxwell's equations, and in particular to the discretized curl operations, **M**, in order to obtain the exact location of the surface currents as discrete delta functions on the Yee grid. This procedure is straightforward, and variations of the derivation can be found in textbook discussions of the total-field/scattered-field approach [1]. Moreover, even if the incident field $\psi_+$ solves the *exact* Maxwell's equations, it will not solve the *discretized* Maxwell's equations without modification. For the case of an incident plane wave in free space, the exact discretized solution for $\psi_+$ can be found by substituting a plane wave into FDTD's Yee discretization, and then solving for the dispersion relation, $\omega(k)$, and the corresponding $E$ and $H$ fields (which are not quite perpendicular to $k$ due to numerical anisotropy) [12]. An efficient recent alternative, described in detail in Chapter 3, accomplishes this "on the fly" as FDTD time-stepping progresses.

In contrast, obtaining $\psi_+$ for geometries like dielectric waveguides must, in general, be done numerically, especially in three dimensions. Suppose one uses a $\psi_+$ solution from an eigenmode solver that is sufficiently converged as to be considered exact. In this case, small wave-source errors, which vanish with increasing FDTD grid resolution, are introduced in FDTD because of the mismatch between the exact $\psi_+$ and the FDTD-discretized $\psi_+$ solutions. This is precisely what was observed in the modeling results shown in Fig. 4.2, where the eigenmode solver used the free plane-wave expansion code, MPB [13], at a high resolution, followed by a simple interpolation onto $\partial\Omega$, the line-segment waveguide cross-section in the FDTD grid. As a result, a small backwards (leftward-propagating) wave artifact was generated within the FDTD waveguide model by the eigenmode current source, carrying approximately $10^{-5}$ of the power of the desired rightward-propagating mode for an FDTD grid resolution of about 27 pixels per wavelength. To eliminate this artifact, one could use the "bootstrap" technique described in [1] where a preliminary FDTD model of the infinitely long waveguide is run to obtain $\psi_+$, using exactly the same spatial and temporal resolutions as the subsequent "working" FDTD run. Of course, this would increase the required computational resources, but once the discretized $\psi_+$ is obtained and stored, it could be used in all subsequent models of the waveguide in question.

A second source of error, which is more difficult to deal with in practice, is dispersion: any nontrivial frequency dependence of the incident-wave solution, $\psi_+$. For a plane-wave $\sim \exp(i\mathbf{k}\cdot\mathbf{x} - i\omega t)$, the exact vacuum dispersion relation is simply $\omega = c|\mathbf{k}|$. However, much more complicated $\omega(k)$ dispersion relations can arise. First, there are numerical dispersion artifacts due to the FDTD spatial and temporal discretizations [1]. Second, a variety of physical dispersions are caused by frequency-dependent material dielectric properties and structural geometric effects such a waveguide cutoff phenomena. In general, suppose that we have solved for the desired incident fields (e.g., plane waves or waveguide modes), $\hat{\psi}_+(\mathbf{x}, \omega)$, at *each* frequency, with a *frequency-independent* normalization (e.g., normalized to unit input power). Furthermore, suppose that we wish to inject a wave *pulse* into our FDTD simulation, with some pulse profile, $p(t)$ (e.g., a Gaussian pulse), and denote its Fourier transform by $\hat{p}(\omega)$. It follows that the desired equivalent currents are given by the following Fourier transform pair:

$$\hat{\boldsymbol{\xi}}(\boldsymbol{x}, \omega) = \delta(\partial\Omega) \begin{bmatrix} \boldsymbol{n}\times \\ -\boldsymbol{n}\times \end{bmatrix} \hat{\boldsymbol{\psi}}_+(\boldsymbol{x}, \omega)\, \hat{p}(\omega) \quad (4.5a)$$

$$\boldsymbol{\xi}(\boldsymbol{x}, t) = \delta(\partial\Omega) \begin{bmatrix} \boldsymbol{n}\times \\ -\boldsymbol{n}\times \end{bmatrix} \boldsymbol{\psi}_+(\boldsymbol{x}, t) * p(t) \quad (4.5b)$$

where $*$ denotes a convolution of the time-domain fields, $\boldsymbol{\psi}_+$ (the inverse Fourier transform of $\hat{\boldsymbol{\psi}}_+$), with the pulse shape $p(t)$. In any time-domain method such as FDTD, of course, we need the time-domain currents, $\boldsymbol{\xi}$. The difficulty is that these convolutions (a *different* convolution at every point $\boldsymbol{x}$ on the surface, $\partial\Omega$) can be cumbersome to perform in general. Several options are:

- Precompute $\boldsymbol{\xi}(\boldsymbol{x}, t)$ via inverse Fourier transformation: Compute $\hat{\boldsymbol{\psi}}_+(\boldsymbol{x}, \omega)$ and $\hat{p}(\omega)$ at a set of discrete $\omega$ (assuming a bandlimited pulse); multiply them; and then perform an inverse fast Fourier transform (FFT) [12, 14]. Unfortunately, this can require a large amount of storage if $\boldsymbol{\xi}(\boldsymbol{x}, t)$ is nonzero over a large surface, $\partial\Omega$.

- Precompute $\boldsymbol{\xi}(\boldsymbol{x}, t)$ via FDTD "bootstrapping" [1]: If $\boldsymbol{\psi}_+(\boldsymbol{x}, t)$ is the field generated by a known current source (e.g., a point source in an infinite waveguide) lying outside $\Omega$, precompute these fields by an FDTD simulation in the incident medium $\boldsymbol{\chi}_+$ (i.e., with no scatterers, etc., using PML to absorb outgoing waves); store their values on $\partial\Omega$; and then convolve them if needed with any auxiliary pulse, $p(t)$, to obtain $\boldsymbol{\xi}(\boldsymbol{x}, t)$. An analogous approach can also be used to convert hard sources into equivalent transparent currents [15]. Again, this can be storage intensive.

- As first reported in [16] and described in detail in Chapter 3, for the special case of an incident plane wave in free space, the computation of $\boldsymbol{\xi}(\boldsymbol{x}, t)$ can be reduced to a *one-dimensional* (1-D) FDTD problem that is co-evolved with the main simulation along the incident wavevector, $\boldsymbol{k}$. This greatly reduces the storage requirements because all points in the same phase plane are redundant.

- Given a bandwidth of interest (since $|\hat{p}|$ is typically small outside some bandwidth), one could apply standard filter-design techniques from digital signal processing [17] to approximate $\hat{\boldsymbol{\psi}}_+(\boldsymbol{x}, \omega)$ in that bandwidth by a simple rational function of $e^{j\omega}$. This would represent a "finite impulse-response" (FIR) or recursive "infinite impulse-response" (IIR) filter with a small amount of storage (the filter "tap coefficients") for each $\boldsymbol{x}$. In this way, the $\boldsymbol{\psi}_+ * p$ convolution could be computed during the FDTD simulation with minimal storage per $\boldsymbol{x}$ (especially if $\hat{\boldsymbol{\psi}}_+$ is a slowly varying function of $\omega$ that is well approximated by a low-degree polynomial in the desired bandwidth), at the expense of greater software complexity [18].

In the absence of one of these techniques, a simple work-around is given in the next section.

## 4.3  SEPARATING INCIDENT AND SCATTERED FIELDS

In simulations where one is computing the scattered, reflected, or similar fields, it is necessary to distinguish the scattered fields from the incident fields. If the equivalent currents, $\boldsymbol{\xi}$, from the previous section are applied exactly, taking into account all discretization and dispersion effects, then the currents produce *zero* fields outside Ω (to nearly machine precision), so that any fields outside Ω are *only* the scattered fields. However, as noted above, exactly accounting for discretization and dispersion effects can be cumbersome. Moreover, in many cases, it is possible and convenient to use a much simpler method than the equivalent-currents prescription:

- In a single-mode waveguide as in Fig. 4.2(a), a point-dipole (or line-segment) current source is sufficient to excite the waveguide mode, as long as the source is far enough away from the region of interest that any other (non-guided) fields radiate away.

- To create an incident plane wave on a *periodic* surface, as described in Section 4.5, only the symmetry of the source matters. Any planar current source with the correct $e^{i\mathbf{k}\cdot\mathbf{x}}$ phase relationship and the desired polarization produces a plane wave, but generally traveling in *both* directions away from the source.

It turns out that there is a very simple and efficient way to exactly separate the incident and scattered fields, even for sources that are not the exact equivalent currents of a total-field/scattered-field approach: we simply run two FDTD simulations and subtract them. More precisely, to avoid large storage requirements, we subtract the Fourier transforms of the fields, as discussed next.

A typical use of FDTD simulations is to inject an incident electromagnetic pulse into a problem geometry; Fourier-transform the desired response (e.g., the transmitted, reflected, or scattered fields); compute the corresponding power or energy at each frequency; and thereby obtain the entire *spectrum* of the response (e.g., a transmission or absorption spectrum) in a single simulation. That is, for a field, $f(t)$, in response to the exciting pulse, one computes the Fourier transform, $\hat{f}(\omega)$, approximated by a discrete-time Fourier transform (DTFT) of the discrete-time field, $f(n\Delta t)$:

$$\hat{f}(\omega) \;=\; \frac{1}{\sqrt{2\pi}}\int_{-\infty}^{\infty} f(t)e^{i\omega t}dt \;\approx\; \frac{\Delta t}{\sqrt{2\pi}}\sum_{n=-\infty}^{\infty} f(n\Delta t)e^{i\omega n\Delta t} \qquad (4.6)$$

where, for a pulsed field, the sum over $n$ can be truncated when $|f|$ becomes sufficiently small. This computation would be performed for a set of frequencies covering the desired bandwidth. In practice, it is more storage efficient to accumulate the $\hat{f}$ summations as the FDTD simulation progresses, rather than storing $f(n\Delta t)$ for all $n$ and computing $\hat{f}$ in postprocessing, especially if the fields at many spatial points are required [1]. For example, to compute the flux through some surface $S$, one would apply this procedure to obtain $\hat{\mathbf{E}}(\mathbf{x},\omega)$ and $\hat{\mathbf{H}}(\mathbf{x},\omega)$ for $\mathbf{x} \in S$, and then compute the flux spectrum, $P(\omega) = \frac{1}{2}\mathrm{Re}\iint_S (\hat{\mathbf{E}}^* \times \hat{\mathbf{H}})\cdot d\mathbf{S}$.

The procedure to separate the incident and scattered fields is then straightforward. First, perform a simulation with the desired current sources, $\xi$, in the incident medium, $\chi_+$, and compute the Fourier-transformed incident field, $\hat{\psi}_+$, on the flux surface, $S$ (storage proportional to the number of points on $S$ multiplied by the number of desired frequencies). Repeat the calculation with the full medium, $\chi$, including any scatterers or other devices, to obtain the Fourier-transformed *total* fields, $\hat{\psi}$, and then subtract them to obtain the Fourier-transformed scattered fields, $\hat{\psi}_- = \hat{\psi} - \hat{\psi}_+$. One can then compute the scattered power spectrum and so on as desired. As an additional benefit of this procedure, one can use the first $\psi_+$ simulation to compute the exact incident power for normalization purposes.

Figure 4.3 illustrates an example of this procedure to compute the transmission and reflection of a single-mode dielectric waveguide through a 90° bend. Using a current source located a distance, $L$, from the bend, we excite a waveguide mode that is incident on the bend.

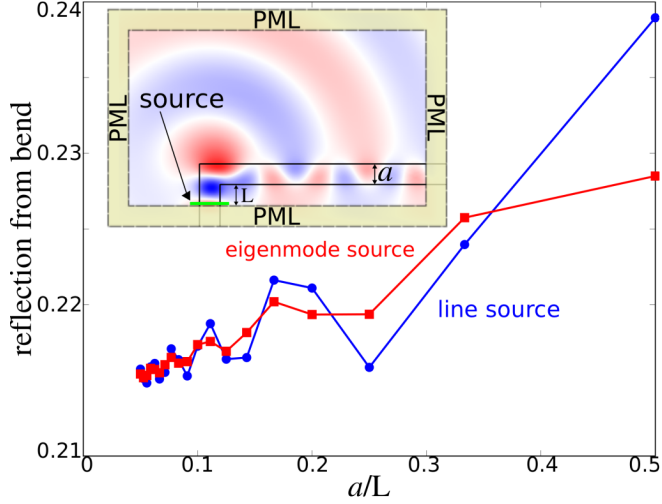

**Fig. 4.3**  FDTD-computed reflection coefficient in the 2-D dielectric waveguide of Fig. 4.2 ($\varepsilon/\varepsilon_0 = 12$, width $a$) with a 90° bend, at the single-mode frequency from Fig. 4.2(a), plotted vs. $a/L$ (the normalized inverse of the distance between the source and the bend). Line plot with square dots: eigenmode-derived $J$ source that excites only the propagating waveguide mode. Line plot with round dots: constant-amplitude $J$ source along a transverse cross-section of the waveguide that excites both the waveguide mode and radiative fields. Inset: visualization of the steady-state $E_z$ field, showing a large radiative-loss component at the bend. The reflection coefficient includes an error term which scales $\sim 1/L$ (hence approaching a straight line in this plot). This is due to the backscattered radiative field contributing to the measured backwards Poynting flux in the waveguide.

We compute the Fourier transform of the resulting field at two places: $\hat{\psi}^{(1)}$ at a distance, $L$, before the bend (at the source location); and $\hat{\psi}^{(2)}$ at a distance, $L$, after the bend. We also perform a second simulation with a *straight* waveguide, and compute the field Fourier transforms, $\hat{\psi}_0^{(1)}$, at the source location, and $\hat{\psi}_0^{(2)}$, after a distance, $2L$. $\hat{\psi}_0^{(2)}$ is used to obtain the incident power, $P_+(\omega)$, and $\hat{\psi}^{(2)}$ is used to obtain the transmitted power, $P_T(\omega)$. Hence, the normalized transmission is $T(\omega) = P_T(\omega)/P_+(\omega)$. $\hat{\psi}_- = \hat{\psi}^{(1)} - \hat{\psi}_0^{(1)}$ is the reflected field, which is used to obtain the reflected power, $P_R(\omega)$. From this, we obtain the normalized reflection, $R(\omega) = P_R(\omega)/P_+(\omega)$, and also the radiative bend loss, $1 - T(\omega) - R(\omega)$.

The key here is to understand the role of the distance, $L$, between the source and the bend. To this end, Figure 4.3 plots the normalized reflection, $R$, versus $a/L$ at the excitation frequency. For comparison, we consider two sources: an eigenmode-derived $J$ source that excites only the propagating waveguide mode, and a constant-amplitude $J$ source along a transverse cross-section of the waveguide that excites both the waveguide mode and radiative fields. (Because of the subtraction procedure, the fact that the latter source excites waveguide modes traveling in both directions is irrelevant.) In Fig. 4.3, the observed $L$ variation of $R$ is due to two factors. First, the simple constant-amplitude $J$ source (but not the eigenmode-derived $J$ source) generates some radiating waves in addition to the guided mode, and a portion of these radiating waves backscatter from the waveguide bend and contribute to $\hat{\psi}^{(1)}$ for any finite $L$. Second, for *both* current sources, when the guided mode hits the waveguide bend, it scatters into radiating fields *in addition* to a backscattered guided mode, and a portion of this scattered radiation contributes to $\hat{\psi}^{(1)}$ for a finite $L$.

However, at the observation point located a distance, $L$, before the waveguide bend, at which the reflection from the bend is computed, the scattering and backscattering radiative effects associated with the bend decrease as $1/L$ in 2-D (and $1/L^2$ in 3-D), the rate at which the radiative field intensity decreases with distance. Thus, as shown in Fig. 4.3, for both current sources, the computed results for reflection from the bend converge to the same value, and at approximately the same rate, as $L \to \infty$ ($a/L \to 0$). We see that there is no great advantage (at least for a single-mode waveguide) in using an equivalent-current source that generates only the waveguide mode, since this procedure does not reduce the contribution of the scattered and backscattered radiating fields.

We close this section by remarking that it is possible to exploit the orthogonality of modes in order to separate the contribution of the backscattered guided mode from any other radiative scattering, and more generally to compute the power scattered into *each* mode of a multimode waveguide. Suppose that one has derived from the FDTD results the Fourier-transformed scattered fields, $\hat{E}$ and $\hat{H}$, at frequency, $\omega$, on a plane, $A$, perpendicular to the waveguide. Suppose also that one has computed (using a mode solver) the fields $\hat{E}_m$ and $\hat{H}_m$ of a particular waveguide mode, $m$, at the same frequency. If this waveguide mode is normalized to unit power, $\frac{1}{2}\mathrm{Re}\iint_A (\hat{E}_m^* \times \hat{H}_m)\cdot d\mathbf{A} = 1$, then the power, $P_m$, carried in $\hat{E}$ and $\hat{H}$ by *only this mode* is given by:

$$P_m = \left| \frac{1}{4} \iint_A \left( \hat{E}_m^* \times \hat{H} + \hat{E} \times \hat{H}_m^* \right) \cdot d\mathbf{A} \right|^2 \tag{4.7}$$

due to the well-known orthogonality properties of modes at a fixed $\omega$ [19–21].[4] Note that this calculation circumvents the need for any space-inefficient convolution because it is performed entirely in the frequency domain, and one only needs to invoke the mode solver at the frequencies where $\hat{E}$ and $\hat{H}$ are desired. A variation on this orthogonality relationship is available for Bloch modes in periodic waveguides or other periodic systems [21]. Some care is required with the normalization and orthogonality of evanescent modes and modes in lossy materials [21].

## 4.4 CURRENTS AND FIELDS: THE LOCAL DENSITY OF STATES

A key point in understanding the relationship between current sources and the resulting fields is this: *the same current radiates a different amount of power depending on the surrounding geometry*. This well-known fact takes on many forms. In antenna theory, the ratio of radiated power, $P$, to the mean-squared current, $I^2$, is known as the *radiation resistance*, $R = P/I^2$, and its dependence on the surrounding geometry can be illustrated analytically by, for example, the method of images for a dipole current ("antenna") above a conducting ground plane [22]. A dipole source in a hollow metallic waveguide emits a power that diverges as a mode cutoff frequency is approached, an effect that has been explained in terms of the frequency-dependent waveguide impedance [23], but which can also be viewed as an example of a much more general phenomenon: *a Van Hove singularity in the density of states*. More precisely, the power radiated by a point-dipole source turns out to be exactly proportional to a fundamental physical quantity: the *local density of states* (LDOS), a measure of how much the harmonic modes of a system overlap with the source point. Equivalently, we can say that the LDOS is proportional to the radiation resistance of a dipole antenna.

---

[4] The field orthogonality relationships are often written without the complex conjugations [19], and it is easy to become confused on this point. For propagating modes in a lossless waveguide, one can do this because the field components transverse to the plane can be chosen to be purely real [21]. However, including the conjugation is convenient for mode solvers that do not enforce this phase choice. For lossy waveguides, dropping the conjugation is essential because the waveguide eigenproblem is complex-symmetric rather than Hermitian [21]. In this case, however, the unit-power normalization procedure is less clear. For evanescent modes, dropping the conjugation is equivalent to a time-reversal procedure that is required in such cases [21]. For periodic systems (and magneto-optic materials), matters are more complicated, but for lossless propagating modes the conjugation is essential [21].

In this section, we will review the derivation of this relationship, and the definition of the LDOS. The LDOS is of central importance not only for understanding classical dipole sources, but also in many physical phenomena that can be understood semiclassically in terms of dipole currents. For example, the spontaneous emission rate of atoms (key to fluorescence and lasing phenomena) is proportional to the LDOS [24–26], and enhancement of the LDOS by a microcavity is known as a *Purcell effect* [27]. A similar enhancement also occurs for nonlinear optical effects [28]. In understanding complex systems, the LDOS also gives a localized measure of the spectrum of eigenfrequencies present in a system, which is extremely useful in assessing a finite structure (e.g., portions of waveguides or periodic media) in which the concept of a dispersion relation (applicable to infinite waveguides or infinite periodic media) does not directly apply. The relationship of LDOS to the power exerted by a current source also makes it easier to generalize the concept of eigenfrequency spectra to lossy or leaky systems, in which the eigenproblem is non-Hermitian and more difficult to analyze directly. We will review the computation of the LDOS in FDTD by inserting dipole sources and performing appropriate Fourier transforms [29–34], or alternatively by computing resonant modes and their Purcell factors.

### 4.4.1 The Maxwell Eigenproblem and the Density of States

The central question is the relationship of the power radiated by a current source to the harmonic modes (eigenmodes) of Maxwell's equations in a given geometry. To illuminate this relationship, we follow the standard approach of expanding in an orthonormal basis of eigensolutions to Maxwell's equations [26, 31]. Let us begin by summarizing the Maxwell eigenproblem, which is discussed in detail elsewhere [35], considering lossless and dispersionless linear materials, $\varepsilon(x) > 0$ and $\mu(x) > 0$, for simplicity. For time-harmonic fields ($\sim e^{-i\omega t}$) in the absence of current sources, Maxwell's equations become $\nabla \times E = i\omega\mu H$ and $\nabla \times H = -i\omega\varepsilon E$. This can be combined into the linear eigenproblem:

$$\frac{1}{\varepsilon}\nabla \times \frac{1}{\mu}\nabla \times E \triangleq \Theta E = \omega^2 E \qquad (4.8)$$

where we have defined the linear operator, $\Theta = (1/\varepsilon)\nabla \times (1/\mu)\nabla \times$ [35]. For appropriate boundary conditions, this operator is *Hermitian* ($\langle E, \Theta E'\rangle = \langle \Theta E, E'\rangle$ for any $E$ and $E'$) and semidefinite ($\langle E, \Theta E\rangle \geq 0$ for any $E$) under the inner product, $\langle E, E'\rangle = \int E^* \cdot \varepsilon E'$. As a consequence, the solutions of (4.8) have real eigenfrequencies, $\omega^{(n)}$, and orthogonal eigenfields, $E^{(n)}$ [35]. For simplicity, we imagine a problem contained in a finite box so that the eigenfrequencies form a discrete set, $\omega_n$, for $n = 1, 2, \ldots$, taking a limit later on to obtain infinite systems. The eigensolutions, $E^{(n)}$, can be chosen to be purely real-valued, if desired.

The density of states (DOS) is then defined by [36, 37]:

$$\mathrm{DOS}(\omega) = \sum_n \delta\left(\omega - \omega^{(n)}\right) \tag{4.9}$$

so that $\int \mathrm{DOS}(\omega)\,d\omega$ is simply a count of the number of modes in the integration interval. If we normalize the eigenfields such that $\langle \boldsymbol{E}^{(n)}, \boldsymbol{E}^{(n)} \rangle = 1$, then we can define a (per-polarization) *local* density of states by [36, 37][5]:

$$\mathrm{LDOS}_\ell(\boldsymbol{x}, \omega) = \sum_n \delta\left(\omega - \omega^{(n)}\right) \varepsilon(\boldsymbol{x}) \left| E_\ell^{(n)}(\boldsymbol{x}) \right|^2 \tag{4.10}$$

so that

$$\mathrm{DOS}(\omega) = \sum_{\ell=1}^{3} \int \mathrm{LDOS}_\ell(\boldsymbol{x}, \omega)\, d^3\boldsymbol{x} \tag{4.11}$$

That is, $\mathrm{LDOS}_\ell(\boldsymbol{x}, \omega)$ gives a measure of $\mathrm{DOS}(\omega)$ weighted by how much of the energy density of each mode's electric field is at position $\boldsymbol{x}$ in direction $\ell$. (Some applications, modeled by dipoles with random or fluctuating orientations, may instead need the polarization-independent LDOS, which is equal to $\sum_\ell \mathrm{LDOS}$.) Remarkably, we will find that $\mathrm{LDOS}_\ell(\boldsymbol{x}, \omega)$ is exactly proportional to the power radiated by an $\ell$-oriented point-dipole current located at $\boldsymbol{x}$. In fact, that radiated-power (Green's function) version of the LDOS formula is often regarded as the more fundamental (albeit less obvious) form of the definition because it is more easily generalized to systems with loss.

Sometimes, authors say "DOS" when they are really referring to an LDOS. For an infinite system with translational invariance (or periodicity), the DOS is usually reported per unit length (or per period) in each direction of invariance (or periodicity). In a homogeneous medium (where the LDOS is position invariant), the DOS per unit distance in each direction is then precisely the LDOS.

### 4.4.2 Radiated Power and the Harmonic Modes

For any time-harmonic electric current source, $\boldsymbol{J}(\boldsymbol{x})e^{-i\omega t}$, we can combine Maxwell's equations, $\nabla \times \boldsymbol{E} = i\omega\mu\boldsymbol{H}$ and $\nabla \times \boldsymbol{H} = -i\omega\varepsilon\boldsymbol{E} + \boldsymbol{J}$, to obtain the equation:

$$\left( \nabla \times \frac{1}{\mu} \nabla \times\ -\omega^2 \varepsilon \right) \boldsymbol{E} = i\omega \boldsymbol{J} \tag{4.12}$$

---

[5]This is not the only way to define an LDOS. For example, one could define an LDOS weighted by the magnetic field energy, $\mu|\boldsymbol{H}^{(n)}|^2$ (normalized to $\int \mu|\boldsymbol{H}^{(n)}|^2 = 1$), instead of the electric field, which would be proportional to the power radiated by a *magnetic* dipole current. However, because so many physical phenomena are related to electric dipole sources, the electric-field LDOS seems to be more commonly used. Alternatively, in some circumstances (e.g., in considering the total energy of thermal fluctuations), it is useful to have an LDOS weighted by the total electromagnetic energy density, $\frac{1}{2}\varepsilon\left|\boldsymbol{E}^{(n)}(\boldsymbol{x}_0)\right|^2 + \frac{1}{2}\mu\left|\boldsymbol{H}^{(n)}(\boldsymbol{x}_0)\right|^2$ [38], which corresponds to an average of electric and magnetic dipole powers.

Equivalently, $\boldsymbol{E} = i\omega(\Theta - \omega^2)^{-1}\varepsilon^{-1}\boldsymbol{J}$ in terms of $\Theta$ from Section 4.4.1. Applying Poynting's theorem [39], the total radiated power, $P$, plus absorbed power in a lossy time-harmonic system, is equal and opposite to the time-average work done by the electric field on the electric current:

$$P = -\frac{1}{2}\mathrm{Re}\int \boldsymbol{E}^* \cdot \boldsymbol{J}\, d^3x = -\frac{1}{2}\mathrm{Re}\langle \boldsymbol{E}, \varepsilon^{-1}\boldsymbol{J}\rangle \tag{4.13}$$

We can relate $P$ to the spectrum of eigenfrequencies, $\omega^{(n)}$, by expanding $\varepsilon^{-1}\boldsymbol{J}$ in the basis of $\boldsymbol{E}^{(n)}$, using the orthogonality of the modes:

$$\varepsilon^{-1}\boldsymbol{J} = \sum_n \boldsymbol{E}^{(n)}\langle \boldsymbol{E}^{(n)}, \varepsilon^{-1}\boldsymbol{J}\rangle \tag{4.14}$$

Now, we can substitute the eigenequation, $\Theta \boldsymbol{E}^{(n)} = (\omega^{(n)})^2 \boldsymbol{E}^{(n)}$, into $\boldsymbol{E} = i\omega(\Theta - \omega^2)^{-1}\varepsilon^{-1}\boldsymbol{J}$ to obtain:

$$\boldsymbol{E}(\boldsymbol{x}) = \sum_n \frac{i\omega \boldsymbol{E}^{(n)}(\boldsymbol{x})\langle \boldsymbol{E}^{(n)}, \varepsilon^{-1}\boldsymbol{J}\rangle}{(\omega^{(n)})^2 - \omega^2} \tag{4.15}$$

However, some care is required to substitute this expression into $P$. On the one hand, the numerator of $\langle \boldsymbol{E}, \varepsilon^{-1}\boldsymbol{J}\rangle$ is then $i\omega|\langle \boldsymbol{E}^{(n)}, \varepsilon^{-1}\boldsymbol{J}\rangle|^2$, which is purely imaginary, making it seem naively as if we will obtain $-\frac{1}{2}\mathrm{Re}\langle \boldsymbol{E}, \varepsilon^{-1}\boldsymbol{J}\rangle = 0$. On the other hand, this $\boldsymbol{E}(\boldsymbol{x})$ expression diverges when $\omega = \omega^{(n)}$, the physical result of driving a lossless oscillator at resonance for an infinite time. A basic problem here is that the equations are singular at $\omega = \omega^{(n)}$. At this frequency, the solutions are not unique since we can add any multiple of $\boldsymbol{E}^{(n)}$. Some careful regularization is required in order to treat this singularity properly to obtain the delta functions in the LDOS. In particular, the correct approach is to consider a *lossy* system (complex $\varepsilon$), for which the equations are nonsingular at all real $\omega$, and then to take the *limit* as the losses go to zero (Im $\varepsilon \to 0^+$). In effect, add an "infinitesimal loss." Physically, every material except vacuum has some loss. In an infinite system, adding an infinitesimal loss in this manner is equivalent to applying a radiation boundary condition to obtain a unique solution, since loss eliminates incoming waves from infinity. Adding an infinitesimal loss is also sometimes thought of as "enforcing causality" [26], which requires the Green's function to be analytic in the upper-half complex-$\omega$ plane [39].

Although this infinitesimal loss is often written formally as eigenfrequencies, "$\omega^{(n)} - i0^+$" or similar [24, 36], it is useful to go through this limiting process step by step [26, 31]. Suppose that we add a small positive imaginary part to $\varepsilon$, corresponding to an absorption loss. The operator, $\Theta$, is no longer Hermitian, but is still complex symmetric under the unconjugated "inner product," $\langle \boldsymbol{E}, \boldsymbol{E}'\rangle = \int \boldsymbol{E} \cdot \varepsilon \boldsymbol{E}'$. Orthogonality of modes still follows, but we now have complex eigenfrequencies, $\omega_c^{(n)} = \omega^{(n)} - i\gamma^{(n)}$, with $\gamma^{(n)} > 0$ (exponential decay). (For a small Im $\varepsilon$, one can show from perturbation theory that the real part of the frequency, $\omega^{(n)}$, is unchanged to first order, compared to the lossless system [35].) Now, substituting the above equations into $P$, we find:

$$P = \frac{\omega}{2}\mathrm{Im}\sum_n \frac{\langle \boldsymbol{E}^{(n)}, \varepsilon^{-1}\boldsymbol{J}\rangle \langle \boldsymbol{E}^{(n)}, \varepsilon^{-1}\boldsymbol{J}^*\rangle}{(\omega^{(n)} - i\gamma^{(n)})^2 - \omega^2} \tag{4.16}$$

Now, suppose that the loss is small, so that $\gamma^{(n)} \ll \omega^{(n)}$ for all modes. For this case, we note that $\gamma^{(n)}/\omega^{(n)}$ is proportional to $\operatorname{Im}\varepsilon/\operatorname{Re}\varepsilon$ [35], and $\boldsymbol{E}^{(n)}$ is approximately the real (lossless) eigenmode plus a correction of order $\operatorname{Im}\varepsilon/\operatorname{Re}\varepsilon$. Things simplify even further because the contribution of each term is negligible except when $\omega$ is close to $\omega^{(n)}$, making the denominator approximately 0. Altogether, to lowest order in $\operatorname{Im}\varepsilon/\operatorname{Re}\varepsilon$, one obtains:

$$P \approx \frac{1}{4}\sum_n \frac{\gamma^{(n)}\left|\left\langle \boldsymbol{E}^{(n)}, \varepsilon^{-1}\boldsymbol{J}\right\rangle\right|^2}{(\omega-\omega^{(n)})^2+(\gamma^{(n)})^2} \qquad (4.17)$$

which is simply a sum of *Lorentzian peaks* for each lossy mode. That is, each lossy mode contributes a Lorentzian peak to the radiated power that is proportional to the overlap integral, $\left|\int \boldsymbol{J}^*\cdot\boldsymbol{E}^{(n)}\right|^2$, of the current with that mode. This approximation becomes exact in the limit $\operatorname{Im}\varepsilon\to 0^+$. Importantly, in this $\gamma\to 0$ limit, the Lorentzians approach delta functions: $\lim_{\gamma\to 0}\gamma/(\Delta^2+\gamma^2) = \pi\delta(\Delta)$. One therefore obtains, in the lossless limit:

$$P = \frac{\pi}{4}\sum_n \left|\left\langle \boldsymbol{E}^{(n)}, \varepsilon^{-1}\boldsymbol{J}\right\rangle\right|^2 \delta(\omega-\omega^{(n)}) \qquad (4.18)$$

In some treatments, this whole limiting process is circumvented by employing the complex identity, $\operatorname{Im}[1/(\Delta-i0^+)]=\pi\delta(\Delta)$ [24, 36]. However, the finite-loss result that lossy modes contribute Lorentzian peaks is useful in its own right. We will return to this result in Section 4.4.6 to discuss Purcell enhancement by resonant cavities.

The fact that the power in a lossless system exhibits a delta function peak at each eigenfrequency, $\omega^{(n)}$, assuming nonzero overlap of $\boldsymbol{E}^{(n)}$ with $\boldsymbol{J}$, requires some care to interpret physically. Since the delta function is a distribution rather than a classical function, it is not really valid to evaluate a delta function "at" $\omega=\omega^{(n)}$ to obtain "infinity." Instead, one should always consider a current consisting of a continuous spread of frequencies (a "test function" in distribution-theory parlance [40]). For example, consider a current source, $\boldsymbol{J}(\boldsymbol{x})p(t)$, where $p(t)$ is a pulse source with a continuous Fourier transform, $\hat{p}(\omega)$, representing its frequency spectrum. In computing the *time-integrated* power (i.e., the total work by the current), we can apply Parseval's theorem:

$$\int\left|\left\langle \boldsymbol{E}^{(n)}, \varepsilon^{-1}\boldsymbol{J}p\right\rangle\right|^2 dt = \left|\left\langle \boldsymbol{E}^{(n)}, \varepsilon^{-1}\boldsymbol{J}\right\rangle\right|^2 \int|p(t)|^2 dt = \left|\left\langle \boldsymbol{E}^{(n)}, \varepsilon^{-1}\boldsymbol{J}\right\rangle\right|^2 \int|\hat{p}(\omega)|^2 d\omega \qquad (4.19)$$

Equation (4.18) becomes:

$$\int P\,dt = \frac{\pi}{4}\sum_n \left|\left\langle \boldsymbol{E}^{(n)}, \varepsilon^{-1}\boldsymbol{J}p\right\rangle\right|^2 \left|\hat{p}(\omega^{(n)})\right|^2 \qquad (4.20)$$

This is a sum of *finite* contributions from each mode weighted by the Fourier amplitudes, $\left|\hat{p}(\omega^{(n)})\right|^2$, at the mode frequencies. Physically, injecting a pulse source into a lossless resonant system does a *finite* amount of work, leaving some superposition of resonant modes oscillating losslessly after the pulse source has returned to zero. Our summation gives the energy deposited by the pulse into each mode. (Another subtlety arises in an infinite system, because the number of modes increases with system size, but the overlap of each mode with any localized $\boldsymbol{J}$ decreases, so that in the limit of an infinite system, a current source still expends finite power.)

### 4.4.3 Radiated Power and the LDOS

Given (4.18), deriving the connection between radiated power and the LDOS is straightforward. We merely consider the case of a dipole current source, $\boldsymbol{J}(\boldsymbol{x}) = \boldsymbol{e}_\ell \delta(\boldsymbol{x} - \boldsymbol{x}_0)$, located at $\boldsymbol{x}_0$, where $\boldsymbol{e}_\ell$ is the unit vector in the direction, $\ell \in \{1, 2, 3\}$. In this case, $\left|\left\langle \boldsymbol{E}^{(n)}, \varepsilon^{-1}\boldsymbol{J}\right\rangle\right|^2 = \left|E_\ell^{(n)}(\boldsymbol{x}_0)\right|^2$. Comparing (4.10) and (4.18), we immediately see that:

$$\text{LDOS}_\ell(\boldsymbol{x}_0, \omega) = \frac{4}{\pi} \varepsilon(\boldsymbol{x}_0) P_\ell(\boldsymbol{x}_0, \omega) \tag{4.21}$$

where $P_\ell(\boldsymbol{x}_0, \omega)$ is the power radiated by $\boldsymbol{J} = \boldsymbol{e}_\ell \delta(\boldsymbol{x} - \boldsymbol{x}_0)e^{-i\omega t}$.[6] Hence, the LDOS is exactly proportional to the power radiated by a dipole source, differing only by a factor of $4\varepsilon/\pi$.

The LDOS is often described as the imaginary part of the diagonal of the Green's function, similar to an analogous formula in quantum mechanics [36]. This refers to a definition of the dyadic "photon" Green's function, $\mathbf{G}_\ell$, as solving $(\omega^2 - \boldsymbol{\Theta})\mathbf{G}_\ell(\boldsymbol{x}, \boldsymbol{x}_0) = \boldsymbol{e}_\ell \delta(\boldsymbol{x} - \boldsymbol{x}_0)$, in which a factor, $-i\omega\varepsilon^{-1}$, is missing from the right-hand side compared to $\boldsymbol{E} = i\omega(\boldsymbol{\Theta} - \omega^2)^{-1}\varepsilon^{-1}\boldsymbol{J}$, so that $\boldsymbol{E} = -i\omega\varepsilon^{-1}\mathbf{G}_\ell$, $P_\ell = -\tfrac{1}{2}\int \text{Re}(\boldsymbol{E} * \boldsymbol{J}) = -\tfrac{1}{2}\omega\varepsilon^{-1}\text{Im}\,G_{\ell\ell}$, and $\text{LDOS}_\ell = -(2\omega/\pi)\text{Im}\,G_{\ell\ell}$, matching LDOS definitions given elsewhere [24, 37].

There is an interesting subtlety in the application of (4.21) to spontaneous emission. It has been argued that the spontaneous emission rate is proportional not to the LDOS, but rather to $\text{LDOS}/\varepsilon$ [24, 41] (the "radiative" LDOS). From (4.21), this implies that the spontaneous emission rate is exactly proportional to the power radiated by a dipole, with *no* $\varepsilon$ factor. This is reminiscent of the semiclassical model of spontaneous emission as energy radiated by a classical dipole source [42], and the exact equivalence to the quantum picture can be demonstrated explicitly [25]. However, as a practical matter, one is most often interested in the *relative* enhancement or suppression of spontaneous emission by one structure relative to another, with the emitting atom embedded in the same material in both cases. In such a comparison, the presence or absence of an $\varepsilon$ factor has no effect.

### 4.4.4 Computation of LDOS in FDTD

Relationship (4.21) between the LDOS and the power radiated by a dipole source makes it easy to compute the LDOS in an FDTD model. One can apply the usual technique of injecting an impulsive point-dipole source, $\boldsymbol{e}_\ell \delta(\boldsymbol{x} - \boldsymbol{x}_0) p(t)$, and accumulating the Fourier transforms, $\hat{E}_\ell(\boldsymbol{x}_0, \omega)$, of the field at $\boldsymbol{x}_0$. This yields the complete LDOS spectrum at $\boldsymbol{x}_0$ in a single calculation [30, 33]. The LDOS is then:

$$\text{LDOS}_\ell(\boldsymbol{x}_0, \omega) = -\frac{2}{\pi} \varepsilon(\boldsymbol{x}_0) \frac{\text{Re}\left[\hat{E}_\ell(\boldsymbol{x}_0, \omega)\,\hat{p}(\omega)^*\right]}{|\hat{p}(\omega)|^2} \tag{4.22}$$

---

[6] There is a subtlety in applying $P = -\tfrac{1}{2}\text{Re}\left\langle \boldsymbol{E}, \varepsilon^{-1}\boldsymbol{J}\right\rangle$ to a delta-function current in 2-D or 3-D, because in this case, $\boldsymbol{E}$ diverges at the location of the current. However, the *real part* of $\boldsymbol{E}$ does *not* diverge for $\boldsymbol{J}(\boldsymbol{x}) = \boldsymbol{e}_\ell \delta(\boldsymbol{x} - \boldsymbol{x}_0)$ in the case of a lossless medium at $\boldsymbol{x}_0$. It is easily verified that $-\tfrac{1}{2}\text{Re}\,E_\ell(\boldsymbol{x}_0)$ is then equal to the radiated power as Poynting's theorem demands (or $\tfrac{1}{2}\omega\,\text{Im}\,E_\ell$ in the common convention where the dipole, $\boldsymbol{J}$, is multiplied by $-i\omega$ [39]). Matters are more subtle for lossy materials at $\boldsymbol{x}_0$, but it has been argued that the discretization in FDTD is equivalent to standard analytical regularizations [34].

Here, we need to normalize by the pulse spectrum, $|\hat{p}(\omega)|^2$, in order to obtain the power exerted by a unit-amplitude dipole (assuming linear materials), equivalent to computing the radiation resistance. Most commonly, we are not interested in the LDOS as an absolute quantity, but rather the relative enhancement (or suppression) of LDOS in one system relative to another — for example, the LDOS in some structure relative to the LDOS in a homogeneous medium. Again, this is straightforward: just compute the LDOS twice, once in the reference system (e.g., vacuum) and once in another geometry, and divide the two. This has the useful side effect of canceling any normalization factors (including $|\hat{p}(\omega)|^2$, but *not* the $\hat{p}^*$ factor inside Re) or choices of units. As will be noted in Section 4.10, when computing the LDOS of a resonant cavity with a long lifetime, it can be more efficient to employ an alternate "Purcell" formula for the LDOS following a resonant-mode calculation, unless more sophisticated Padé extrapolation techniques [43–45] are used to compute the Fourier transform of the slowly decaying fields.

Figure 4.4 illustrates an example of such a calculation. Here, we compute the LDOS spectrum at a point inside a finite *photonic crystal* comprised of a 2-D square lattice of dielectric rods ($\varepsilon = 12\varepsilon_0$) in air. For the assumed out-of-plane point-dipole current source, an infinite periodic system of this type would exhibit "photonic bandgaps," i.e., ranges of $\omega$ in which there are no electromagnetic waves propagating in any direction in the crystal [35]. In such an infinite system, (4.10) tells us that the current source would radiate zero power at any $\omega$ in the bandgaps, since there are no modes, $\omega^{(n)}$, in these gaps to make the delta function nonzero.

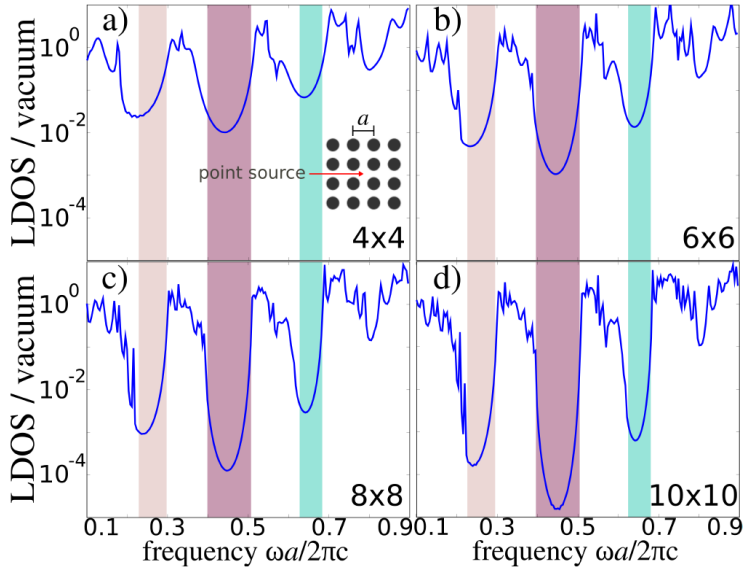

**Fig. 4.4** Radiated power spectrum — proportional to the local density of states (LDOS) — for an out-of-plane point-dipole current source radiating in the center of a finite 2-D photonic crystal (inset: period $= a$ square lattice of dielectric rods of radius $= 0.2a$ and refractive index $= \sqrt{12}$, in air), normalized by a calculation for the same dipole source radiating in free space. The spectrum is drastically changed depending on how many rods (from $4 \times 4$ to $10 \times 10$) surround the source. The LDOS in the photonic bandgaps (shaded regions) decreases exponentially with an increasing number of rods. The LDOS experiences Fabry-Perot oscillations at other frequencies due to the partial reflections at the edges of the photonic crystal, and approaches a discontinuity (Van Hove singularity) at the edges of the bandgaps.

For a finite photonic crystal surrounded by an infinite air region (truncated in FDTD by PML absorbers), the LDOS in the bandgap is nonzero. This is because an evanescent tail of the field produced by the dipole source "leaks" through the photonic crystal and radiates into the air. However, this effect decreases *exponentially* as the crystal size increases. Furthermore, at the edges of the bandgaps, as $\omega$ approaches the cutoff of a band of propagating modes in the crystal, the LDOS becomes more and more singular (here, discontinuous) as the size of the crystal increases. This is called a *Van Hove singularity*, and is discussed in Section 4.4.5.

For the case of spontaneous emission for an atomic transition with a finite Lorentzian linewidth, $\gamma_0$, around a transition frequency, $\omega_0$, the correct enhancement factor is not simply the LDOS at $\omega_0$. In fact, it is an average of the LDOS over the transition linewidth. In particular, the emission rate for such an $\ell$-directed dipole transition at $x_0$ is proportional to [25, 32]:

$$\int \frac{1}{\varepsilon(x_0)} \text{LDOS}_\ell(x_0, \omega) \frac{\gamma_0/\pi}{(\omega-\omega_0)^2 + \gamma_0^2} d\omega \quad (4.23)$$

This converges to $\text{LDOS}_\ell(x_0, \omega_0)/\varepsilon(x_0)$ as $\gamma_0 \to 0$. Therefore, it is reasonable to simply use $\text{LDOS}_\ell(x_0, \omega_0)$ as long as $\gamma_0$ is much smaller than any feature sizes in the LDOS spectrum. Taking the Fourier transform to the time domain, expression (4.23) is equivalent to computing the total work done by a *decaying* dipole source, $J(x,t) \sim e_\ell \delta(x-x_0)\exp(-i\omega_0 t - \gamma_0 t/2)$, for $t > 0$ [32]. However, this is more efficient (i.e., a shorter-time simulation) than postprocessing $\text{LDOS}_\ell(x_0, \omega)$ only if $\gamma_0$ is large (faster decay) compared to the decay rates of the optical modes, and is less efficient if $\gamma_0$ is small.

### 4.4.5 Van Hove Singularities in the LDOS

The LDOS — again, just the power emitted by a dipole source — exhibits singularities in a waveguide or a periodic system whenever a cutoff frequency is approached. These are known (from solid-state physics) as *Van Hove singularities* [46]. These singularities have many important physical consequences. For example, they can be understood as the feedback mechanism in distributed feedback (DFB) lasers. More generally, it is important to understand the relationship between the dispersion characteristic, $\omega(k)$, and the LDOS or DOS, in order to understand the effect of current sources in such structures.

For simplicity, consider a system (e.g., a waveguide) that is invariant in a single direction (say $z$), so that the fields can be written in the separable form, $E_k(x,y)\exp[i(kz-\omega t)]$, with some modal dispersion relation(s), $\omega_n(k)$. To understand the DOS of this system, start with a *finite* system in the $z$-direction of length, $L$, with *periodic* boundary conditions in $z$. In this case, the only allowed solutions are the modes where $k = 2\pi m/L$ is an integer multiple, $m$, of $2\pi/L$. From (4.9), the DOS per unit length, $L$, is then:

$$\frac{\text{DOS}(\omega)}{L} = \frac{1}{L}\sum_n\sum_m \delta[\omega - \omega_n(2\pi m/L)] = \frac{1}{2\pi}\sum_n\left\{\sum_m \delta[\omega - \omega_n(2\pi m/L)]\frac{2\pi}{L}\right\} \quad (4.24)$$

Now, if we take the $L \to \infty$ limit, it is clear that $\sum_m \to \int dk$, with $(2\pi/L) = \Delta k \to dk$. Hence:

$$\frac{\text{DOS}}{\text{per length}}(\omega) = \frac{1}{2\pi}\sum_n \int_{-\infty}^{\infty}\delta[\omega - \omega_n(k)]dk \quad (4.25)$$

Even though the DOS of a *finite* system is "spiky" (a sum of $\delta$ functions), the DOS/length of an infinite system is (mostly) continuous, due to this $\int dk$. A similar argument applies to discrete periodic systems (photonic crystals), in which case one integrates the dispersion relations of the Bloch modes. The LDOS is similar to the DOS, except that, like (4.10), there is an additional factor of $\varepsilon |E_\ell|^2$ weighting the integral.

The only places that singularities arise are at cutoffs (or in general, at points of zero group velocity, $d\omega_n/dk = 0$). For example, suppose that $\omega_1(k)$ has a cutoff at $k = 0$ for a frequency, $\omega_c$ (e.g., for a hollow metal waveguide). Let us apply a Taylor expansion, $\omega_1(k) \approx \omega_c + \alpha k^2$, near $k = 0$, with $\alpha = \frac{1}{2} d^2\omega_1/dk^2|_{k=0} > 0$. In this case, the $\int dk$ in the DOS becomes (near $\omega_c$):

$$\int \delta[\omega - \omega_1(k)] dk \approx \int \delta(\omega - \omega_c - \alpha k^2) dk = \begin{cases} 0 & \omega < \omega_c \\ 0.5[\alpha(\omega - \omega_c)]^{-1/2} & \omega > \omega_c \end{cases} \quad (4.26)$$

with the $(\omega - \omega_c)^{-1/2}$ term arising from a Jacobian factor, $1/2\alpha k$. (Note that the $\delta$ function is only nonzero when $\omega = \omega_c + \alpha k^2$, and hence, $k = [(\omega - \omega_c)/\alpha]^{1/2}$). Thus, the DOS *diverges* with an integrable $(\omega - \omega_c)^{-1/2}$ singularity as $\omega$ approaches the cutoff frequency from above. This is true for any quadratic band edge in *one dimension* of translational symmetry. A similar conclusion is reached for the LDOS, of course. With two dimensions of translational symmetry (e.g., a planar waveguide), one obtains a discontinuity in the DOS (and LDOS) at a band edge; and in three dimensions, there is a continuous $(\omega - \omega_c)^{1/2}$ singularity.[7]

Since the LDOS is proportional to the spontaneous emission rate of excited atoms (semiclassical dipoles), these singularities in the LDOS at band edges (cutoffs) can lead to lasing via "distributed feedback." In an FDTD simulation of a hollow metal waveguide, the $(\omega - \omega_c)^{-1/2}$ singularities lead to a "spectral distortion" of the output of a dipole source, where the dipole source emits more power as the cutoff is approached. This effect was explained by other authors in the language of the waveguide "impedance" [23]. (Of course, in a *finite-length* waveguide surrounded by vacuum, the LDOS is finite — a current radiates finite power — essentially because the zero group-velocity band-edge solutions can escape from the ends of the waveguide.)

### 4.4.6 Resonant Cavities and Purcell Enhancement

A lossless localized mode yields a $\delta$-function spike in the LDOS, whereas a *lossy* localized mode — a *resonant cavity mode* — leads to a Lorentzian peak. This was shown in (4.17) for losses due to a small absorption, but is generically true for any loss mechanism, including "leaky" modes with radiative losses.[8] The large enhancement in the LDOS at the resonant peak is known as a *Purcell effect*, named after Purcell's proposal for enhancing spontaneous emission of an atom in a cavity (by analogy with a microwave antenna resonating in a metal box) [27]. There is a famous formula for this enhancement factor arising from the LDOS derivation above, as will be reviewed next.

---

[7]To understand this dependence on the problem dimension, consider an *isotropic* band-edge shape, $\omega_1(\mathbf{k}) \approx \omega_c + \alpha |\mathbf{k}|^2$. Performing the $\mathbf{k}$ integration in cylindrical or spherical coordinates then yields a $\int 2\pi k\, dk$ or $\int 4\pi k^2 dk$ integral, respectively. The additional $k$ or $k^2$ factors multiply the integration result by the terms, $(\omega - \omega_c)^{1/2}$ or $(\omega - \omega_c)$.

[8]The theory of leaky "modes" is somewhat subtle because the modes are not strictly eigenfunctions [47]. They can be defined as poles in the Green's function that are close to the real-$\omega$ axis (slightly *below* it due to causality), which consequently contribute a Lorentzian peak to $\text{Re}(\mathbf{E}^* \cdot \mathbf{J})$.

Typically, one characterizes the lifetime of a lossy mode by a dimensionless *quality factor*, $Q^{(n)} = \omega^{(n)}/2\gamma^{(n)}$ [35]. In terms of $Q^{(n)}$, the contribution to the LDOS at a resonant peak, $\omega = \omega^{(n)}$, from (4.17) is given by:

$$\frac{1}{\pi}\varepsilon(\boldsymbol{x}_0)\left|E_\ell^{(n)}(\boldsymbol{x}_0)\right|^2 \frac{2Q^{(n)}}{\omega^{(n)}} \tag{4.27}$$

Suppose that we choose $\boldsymbol{x}_0$ to be the point where $\varepsilon|\boldsymbol{E}^{(n)}|^2$ is *maximum*, and we consider the polarization-averaged $\text{LDOS}(\boldsymbol{x}_0) = \sum_\ell \text{LDOS}_\ell(\boldsymbol{x}_0)$ (e.g., for dipoles with random orientation). Then, because we normalized $\int \varepsilon|\boldsymbol{E}^{(n)}|^2 = 1$, the $\varepsilon(\boldsymbol{x}_0)|\boldsymbol{E}^{(n)}(\boldsymbol{x}_0)|^2$ term has units of inverse volume, the *modal volume*, $V^{(n)}$ [27, 35]:

$$V^{(n)} = \frac{\int \varepsilon|\boldsymbol{E}^{(n)}|^2}{\max \varepsilon|\boldsymbol{E}^{(n)}|^2} \tag{4.28}$$

This is essentially the volume in which $\varepsilon|\boldsymbol{E}^{(n)}|^2$ is not small. In this case, the resonant mode's contribution to the LDOS at $\omega^{(n)}$ is given by:

$$\text{resonant LDOS} \approx \frac{2}{\pi\omega^{(n)}}\frac{Q^{(n)}}{V^{(n)}} \tag{4.29}$$

If $Q^{(n)}$ is large enough, all other contributions to the LDOS (from the other modes) are negligible at $\omega^{(n)}$. Hence, for a given frequency $\omega^{(n)}$, we obtain an approximate $Q/V$ enhancement in the LDOS at the point of the peak resonant field. This $Q/V$ enhancement is widely known as a *Purcell factor*. Note that, upon examining (4.23), we see that this enhancement factor only applies to spontaneous emission if the atomic transition linewidth, $\gamma_0$, is much smaller than the microcavity linewidth, $\gamma^{(n)}$. If $\gamma_0 \gg \gamma^{(n)}$, then the contribution from the cavity is reduced by a factor, $\gamma^{(n)}/\gamma_0$.

Figure 4.5 illustrates an example of Purcell enhancement of the LDOS. Here, we consider a 2-D perfect-metallic $a \times a$ cavity of finite wall thickness, $0.1a$. One sidewall is assumed to have a small notch of width, $w$, which allows the cavity modes to escape to the surrounding free-space region, as shown in the inset. In the absence of the notch, the lowest-frequency mode with out-of-plane polarization is $E_z^{(1)} = (4/a^2)\sin(\pi x/a)\sin(\pi y/a)$, with a frequency, $\omega^{(1)} = \sqrt{2}c\pi/a$, and a modal volume, $V^{(1)} = a^2/4$. While the notch slightly perturbs this solution, more importantly, it allows radiation into the surrounding region, yielding a finite $Q$. For $w \ll a$, this radiative escape occurs via an evanescent (sub-cutoff) mode of the channel waveguide formed by the notch. It follows from inspection of the evanescent decay rate, $[(\pi/w)^2 - (\omega^{(1)})^2]^{1/2}/c$, that the lifetime scales asymptotically as $Q^{(1)} \sim e^{\#/w}$ for some coefficient $\#$.

The results of Fig. 4.5 validate both this prediction and the LDOS calculations described in this section. Here, the LDOS at the center of the cavity (the point of peak $|\boldsymbol{E}|$) is computed in two ways. The first is via the exact dipole-power expression of (4.21). The second is via the Purcell approximation, (4.29), where the cavity mode and its lifetime $Q$ are obtained using the filter-diagonalization technique. The latter approach is much more efficient for high $Q$ (small $w$), since one must otherwise run the FDTD simulation for a very long time to directly accumulate the Fourier transform of a slowly decaying mode. In Fig. 4.5, the results of the two approaches agree to within the discretization error. Furthermore, the Purcell $Q/V$ approximation is asymptotically linear on a semilog scale versus $1/w$, as predicted. This verifies the LDOS analyses of this section.

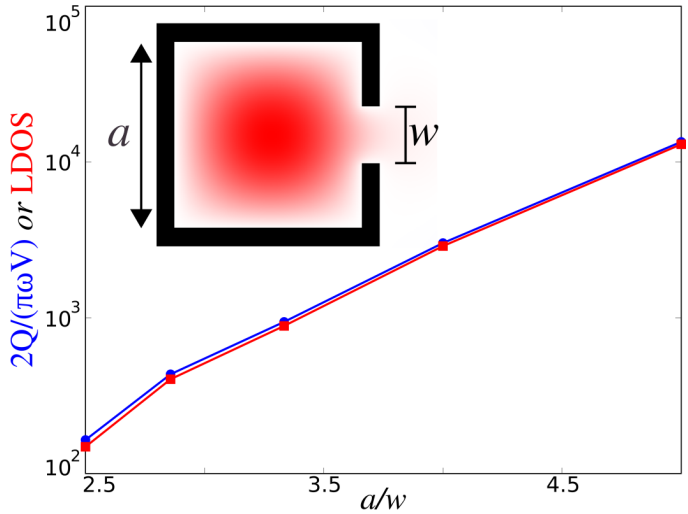

**Fig. 4.5** FDTD-computed Purcell enhancement of the LDOS in a 2-D perfect-metallic $a \times a$ cavity of finite wall thickness, $0.1a$. As shown in the inset, one sidewall of the cavity is assumed to have a small notch of width, $w$, which allows the cavity modes to escape to the surrounding free-space region with some lifetime, $Q$. We consider the fundamental mode with out-of-plane ($E_z$) polarization. Square dots show the LDOS computed by the exact dipole-power formula of (4.21); round dots show the results of the Purcell approximation of (4.29). Agreement of the two approaches is within the computational accuracy.

### 4.5 EFFICIENT FREQUENCY-ANGLE COVERAGE

A common problem in nanophotonics is to evaluate the reflection, transmission, or absorption spectrum of a periodically patterned surface as a function of the incidence angle of light, as depicted schematically in Fig. 4.6. For example, calculating (and optimizing [48]) the absorption of periodically patterned surfaces is of key interest in designing efficient thin-film photovoltaic cells [49]. In this section, we review some basic properties of this problem and describe how it can be efficiently solved in FDTD, including a technique to map out the spectra as a function of *both* frequency and incidence angle by using a minimal number of FDTD simulations with simple pulse (broadband) line-current sources [50 – 52].

Such problems can be solved efficiently in FDTD because they can be reduced to a problem in the *unit cell* of the periodicity (dashed boxes in Fig. 4.6), which is finite in the plane of the surface and can be truncated by PML absorbers in the direction perpendicular to the surface. The key is to impose the correct boundary conditions. Periodicity means that the *structure* is invariant under translation by lattice vectors $\boldsymbol{R}$ [e.g., $\boldsymbol{R} = (a, 0)$ is the primitive lattice vector in the 2-D example of Fig. 4.6(b)]. However, this does *not* mean that the *fields* are periodic or that we can impose periodic boundary conditions. For an incident plane wave $\sim e^{i(\boldsymbol{k} \cdot \boldsymbol{x} - \omega t)}$ with wavevector $\boldsymbol{k}$, the incident fields are *Bloch periodic*:

$$\text{fields}(\boldsymbol{x} + \boldsymbol{R}) = \text{fields}(\boldsymbol{x}) e^{i \boldsymbol{k} \cdot \boldsymbol{R}} \qquad (4.30)$$

That is, the fields are periodic up to a phase shift depending on the components of $\boldsymbol{k}$ in the plane of periodicity.

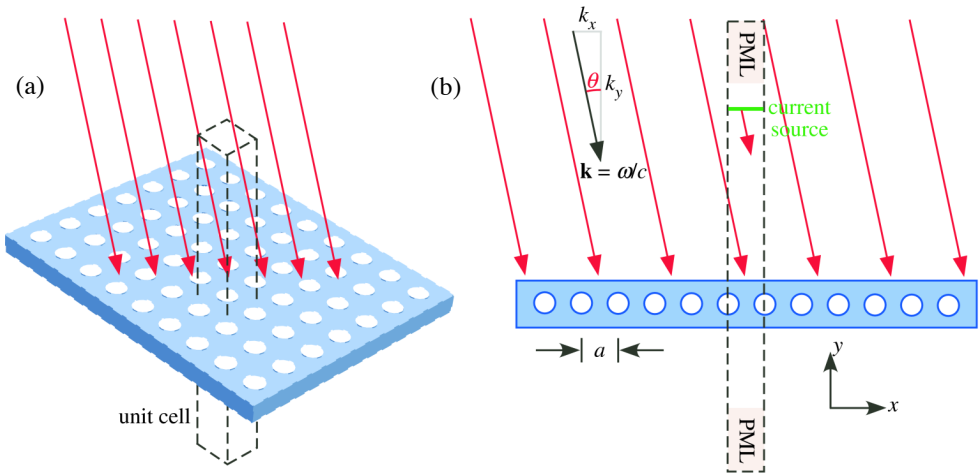

**Fig. 4.6** Plane wave obliquely incident on periodic structures. (a) For a periodically patterned surface in 3-D, the problem can be reduced to a simulation within a unit cell (the dashed parallelepiped) having Bloch-periodic boundaries. (b) Analogous 2-D geometry, a holey waveguide comprised of a dielectric strip perforated by a linear array of holes of period, $a$. Here, the unit cell of the computation is shown as a dashed box, with PML absorbers in the $y$-direction and Bloch-periodic (phase $e^{ik_x a}$) boundary conditions in the $x$-direction, where $k_x$ is the surface-parallel component of the incident wavevector, $k$. In FDTD, the incident wave can be excited with a current source of appropriate phase and polarization extending transversely across the computational cell.

The crucial fact is that, in a periodic system with a Bloch-periodic incident wave, the Bloch periodicity (the surface-parallel component of $k$) is *conserved* for linear materials. The total field (including scattering, etc.) is Bloch periodic with the *same* in-plane $k$ (a kind of conserved "momentum") [35]. This means that we simply need to impose Bloch-periodic boundary conditions in our FDTD simulation in order to reduce the computational domain to the unit cell of the problem. Note that this means that our fields must be simulated as *complex numbers*, which poses no particular difficulty. Because Maxwell equations are unmodified, the real and imaginary parts of the fields only couple via the boundary conditions.

Even though the scattered fields are also Bloch periodic, this does not mean that they are all plane waves with the same $k$. Consider the 2-D problem of Fig. 4.6(b), in which the incident wavevector is $k = (k_x, k_y)$. From (4.30), fields$(x+a)$ = fields$(x) e^{ik_x a}$, so only $k_x$ is "conserved" and not $k_y$. Moreover, even $k_x$ is only conserved up to addition of multiples of $2\pi/a$ ("reciprocal lattice vectors") [35]. For example, a reflected plane wave with $k' = (k_x + 2n\pi/a, k_y')$ is *also* Bloch periodic for any integer, $n$, since $e^{i(2n\pi/a)a} = 1$. These are *diffracted* waves. Summing all such waves for all $n$ yields a Fourier series, and we can equivalently say that the solutions are in general *Bloch waves*: periodic functions of $x$ multiplied by $e^{ik_x x}$ [35]. Furthermore, since the solutions in air must satisfy the dispersion relationship, $\omega = c|k'|$ (frequency is conserved in a linear system), the scattered waves satisfy $k_y' = \pm [(\omega/c)^2 - (k_x + 2n\pi/a)^2]^{1/2}$. For $n = 0$, we obtain $k_y' = \pm k_y$, the "law of equal angles" for the "specular" reflected and transmitted waves. In addition, for any $\omega$, $k_y'$ becomes *imaginary* for sufficiently large $|n|$, corresponding to decaying (evanescent) waves. Hence, there are a finite number of propagating diffracted waves for any finite $\omega$; and for sufficiently small $\omega$, there are *no* diffracted waves except for the specular $n = 0$ waves [35].

Suppose that our incident plane wave is given by fields $\boldsymbol{E}_0 e^{i(\boldsymbol{k}\cdot\boldsymbol{x}-\omega t)}$ and $\boldsymbol{H}_0 e^{i(\boldsymbol{k}\cdot\boldsymbol{x}-\omega t)}$. For a planar current source parallel to the surface [e.g. the source in Fig. 4.6(b)], the equivalent-currents prescription of Section 4.2 would specify the electric and magnetic currents, $\boldsymbol{J} = \boldsymbol{n}\times\boldsymbol{H}_0 e^{i(\boldsymbol{k}\cdot\boldsymbol{x}-\omega t)}\delta(\cdots)$ and $\boldsymbol{K} = -\boldsymbol{n}\times\boldsymbol{E}_0 e^{i(\boldsymbol{k}\cdot\boldsymbol{x}-\omega t)}\delta(\cdots)$. However, one can simplify this in several ways. First, we can use an electric current, $\boldsymbol{J}$, alone, or a magnetic current, $\boldsymbol{K}$, alone. By the equivalence principle, this corresponds to the incident wave plus (or minus) its mirror flip. That is, it generates incident waves propagating both towards and away from the surface, but the latter can be eliminated by the subtraction technique of Section 4.3. Second, it is not necessary to correct for discretization or dispersion effects. It follows from the translational symmetry of the air region (giving, again, conservation of $\boldsymbol{k}$) that an $\sim e^{i\boldsymbol{k}\cdot\boldsymbol{x}}$ current source in air produces a plane wave (in this case, a discrete-space plane wave) with the same in-plane wavevector component. Therefore, all we have to do is insert a planar current source, $\boldsymbol{J}$ or $\boldsymbol{K}$, with the correct in-plane phase relationship and the desired polarization, and perform a second simulation with only air (no surface) for normalization and for subtracting the incident field (or rather, its Fourier transform), as discussed in Section 4.3.

Figure 4.7 illustrates the final step of the analysis: efficiently mapping the reflection or transmission spectrum to its angular response. In this example, the reflection spectrum, $R(\omega,\theta)$, is obtained at multiple angles for the 2-D holey waveguide shown in Fig. 4.6(b).

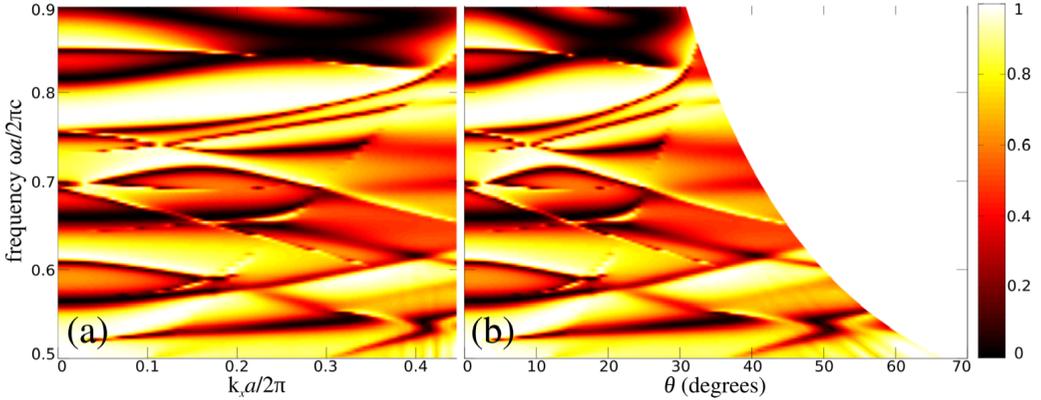

**Fig. 4.7** FDTD-computed reflection spectra, $R$ = reflected power / incident power, for oblique plane-wave incidence on the 2-D holey waveguide of Fig. 4.6(b), characterized by $\varepsilon/\varepsilon_0 = 13$, width = $1.2a$, and hole radius = $0.36a$. (a) $R(\omega, k_x)$, computed for a finite number of fixed $k_x$ values and then interpolated between these vertical cuts to fill the rectangular $\omega \times k_x$ plot; (b) $R(\omega,\theta)$, derived from the data of (a) by first mapping the spectral data computed for each fixed $k_x$ value onto the corresponding curve, $\theta = \sin^{-1}(ck_x/\omega)$, and then interpolating between these curves to fill the remainder of the $\omega \times \theta$ plot.

As usual, we obtain the entire spectrum at once using FDTD by injecting a short (broad-bandwidth) pulse source. We then Fourier transform the result to obtain a flux spectrum normalized by the incident-flux spectrum derived from an empty-space simulation. However, there is one subtlety. When we impose Bloch-periodic boundary conditions in FDTD, we set the fields at one side equal to $e^{ik_x a}$ times the fields at the other side, which gives the *same* $k_x$ for *all* the frequency components, $\omega$. That is, each simulation really computes $R(\omega, k_x)$ for a *fixed* $k_x$.

Note that a fixed $k_x$ does *not* correspond to a fixed angle, $\theta$, except for normal incidence, $k_x = 0$. This is because $\theta$ depends on $\omega$ as well as $k_x$: $\theta(\omega, k_x) = \sin^{-1}(ck_x/\omega)$. Of course, $R(\omega, k_x)$ is useful in its own right, and is plotted for the holey waveguide structure of Fig. 4.6(b) in Fig. 4.7(a). In fact, $R(\omega, k_x)$ is closely related to this structure's dispersion relation (band diagram), $\omega(k_x)$. The leaky modes above the light line, $\omega > ck_x$ [35], yield *Fano resonances*, adjacent peaks and dips in the reflection spectra [53] that can be observed in Fig. 4.7(a).

However, given the FDTD-computed $R(\omega, k_x)$ data, it is straightforward to convert to an $R(\omega, \theta)$ plot by making the change of variables to $\theta(\omega, k_x)$ for each $k_x$. By this procedure, each frequency spectrum at a fixed $k_x$ maps onto its corresponding $\theta(\omega, k_x)$ curve in the $R(\omega, \theta)$ plot. Readily available software is then exercised to interpolate $R(\omega, \theta)$ between these curves, and thereby fill the desired $\omega, \theta$ parameter space. The result is shown in Fig. 4.7(b). Here, note that the range of $\theta$ depends on $\omega$. [Of course, we could obtain a larger range of $\theta$ by going to larger $k_x$ values in the FDTD simulations. We could also display a "normal" looking $R(\omega, \theta)$ plot simply by cropping the $\omega, \theta$ data to a rectangular region.]

## 4.6   SOURCES IN SUPERCELLS

A common problem that arises in FDTD simulations of periodic systems is that FDTD is usually formulated using orthogonal Cartesian grids [1], whereas periodic systems commonly have nonorthogonal (parallelogram/parallelepiped) unit cells [35]. This mismatch can be dealt with in a variety of ways. FDTD can be reformulated to use a nonorthogonal grid that represents the desired unit cell directly [1, 54], but changes to the core FDTD algorithm may require major revision of one's existing FDTD software. Essentially equivalent to this, a coordinate transformation can be used to map the nonorthogonal lattice into an orthogonal lattice. However, this requires appropriately specifying anisotropic materials and transformations of the fields and sources in order to relate them to the original problem [55]. Another option is to use a "staircased" version of the nonorthogonal unit cell with an orthogonal grid [56].

A simple and efficient technique to model a nonorthogonal unit cell in FDTD, without changing the core FDTD algorithm and without changing the materials or fields, is to modify the boundary conditions to employ an orthogonal computational cell. Here, instead of Bloch-periodic, one can use "skewed" Bloch-periodic boundary conditions in which the fields at one side of the computational cell are related to the fields at the other side plus a lateral shift [57, 58].

However, it is tempting to use FDTD *with no modifications whatsoever* in the common case where the nonorthogonal periodic lattice can be described with an orthogonal *supercell*. This is a periodic unit cell that is larger than the primitive unit cell, albeit with an increase in the computational volume. For example, this is true for triangular/hexagonal lattices in 2-D, and fcc/bcc/fct lattices in 3-D [35]. In this section we review the fact that the simplicity of a supercell comes at the price of introducing unwanted additional solutions, due to a *band-folding* phenomenon. Fortunately, there is a simple modification of the *source terms* in FDTD (and, optionally, of the postprocessing as well) that mostly eliminates this problem.

Band folding is easiest to describe in a 1-D example, as illustrated in Fig. 4.8. A typical goal in analyzing a periodic system (e.g., a photonic crystal) is to map out the dispersion relation, $\omega_n(\mathbf{k})$, of the solutions propagating within the crystal, which take the form of *Bloch waves*: $\mathbf{E}, \mathbf{H} \sim$ (periodic function)$e^{i(\mathbf{k}\cdot\mathbf{x} - \omega_n t)}$, i.e., plane waves multiplied by a periodic envelope [35]. In 1-D with period $a$, this means that the field components are of the form $p_k(x)e^{ikx}$, where $p_k(x+a) = p_k(x)$. An example of such bands, $\omega_n(k)$, for $n = 1, 2, 3$ is shown for a simple multilayer film in Fig. 4.8.

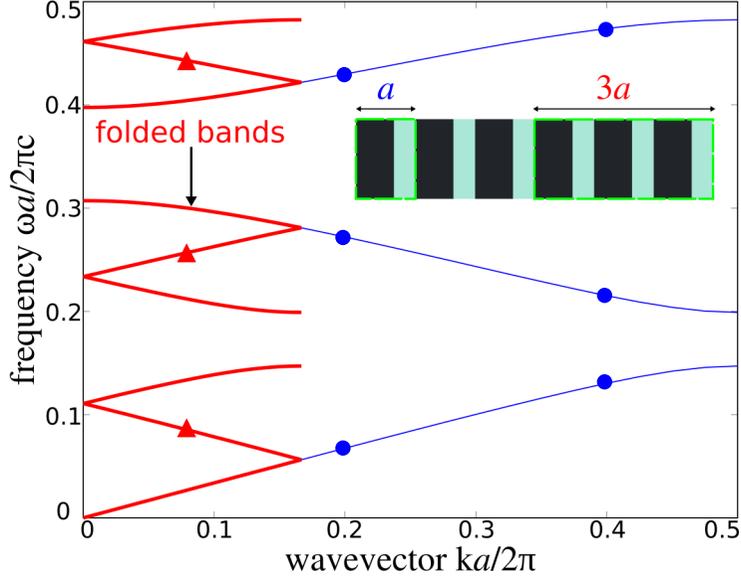

**Fig. 4.8** 1-D example of band folding within a supercell for a multilayer film of period = $a$, alternating between $\varepsilon/\varepsilon_0 = 1$ (thickness = $0.3a$) and $\varepsilon/\varepsilon_0 = 12$ (thickness = $0.7a$) [35] (inset). The curves of the dispersion relation, $\omega_n(k)$, are marked by round dots. There are three bands in this frequency range, separated by photonic bandgaps in which no propagating solutions exist. However, if we solve the same problem in a supercell of size, $3a$, the solutions are the bands "folded" three times (curves marked by triangular dots). These are the same solutions, but the meaning of the $k$ label has changed.

In Fig. 4.8, we only plot for $k \in [0, \pi/a]$ the *irreducible Brillouin zone* [35]. Here, $\omega_n(-k) = \omega_n(+k)$ by either mirror symmetry or time-reversal symmetry. Furthermore, $\omega_n(k+2\pi/a) = \omega_n(k)$ because $p_{k+2\pi/a}(x)e^{i(k+2\pi/a)x} = [p_{k+2\pi/a}(x)e^{i2\pi x/a}]e^{ikx} = [\text{periodic}]e^{ikx}$ is a Bloch solution at $k$ as well as at $k+2\pi/a$. Now, suppose that we solve the system using a three-period ($3a$) supercell. This obviously describes the same structure, and hence, would seem to imply the same solutions. In fact, we do get the same solutions in a sense, but they are mixed up because the meaning of $k$ has changed! That is, suppose we impose Bloch-periodic boundary conditions, so that we are asking for field solutions of the form $f(x)$ such that $f(x+3a) = e^{ik(3a)}f(x)$. This is equivalent to requiring $f(x) = \tilde{p}(x)e^{ikx}$, where $\tilde{p}(x+3a) = \tilde{p}(x)$ is periodic with the period, $3a$. If we compare to our original solution, we find that $p_k(x)e^{ikx}$ is indeed a solution with the new boundary conditions. The problem is that we now get *new* solutions $p_{k\pm 2\pi/3a}(x)e^{i(k\pm 2\pi/3a)x}$ from $k\pm 2\pi/3a$, since $e^{\pm i(2\pi/3a)3a} = 1$. This is shown by the curves marked by the triangles in Fig. 4.8, where the original bands are "folded" onto the new irreducible Brillouin zone, $[0, \pi/3a]$, of the supercell [35]. Of course, these still solve Maxwell's equations, but understanding the dispersion relation has been complicated.

This band-folding effect is even more difficult to disentangle for supercells in 2-D or 3-D. For example, consider the case of a triangular lattice (period $a$) of dielectric rods in air, shown in Fig. 4.9(a). This geometry admits a rectangular $a \times \sqrt{3}a$ supercell that has twice the area of the rhombus unit cell. We wish to compute the bands, $\omega_n(\mathbf{k})$, around the boundaries ($\Gamma$–M–K) of its irreducible Brillouin zone of Fig. 4.9(b) [35], compared to an alternate, plane-wave method [13].

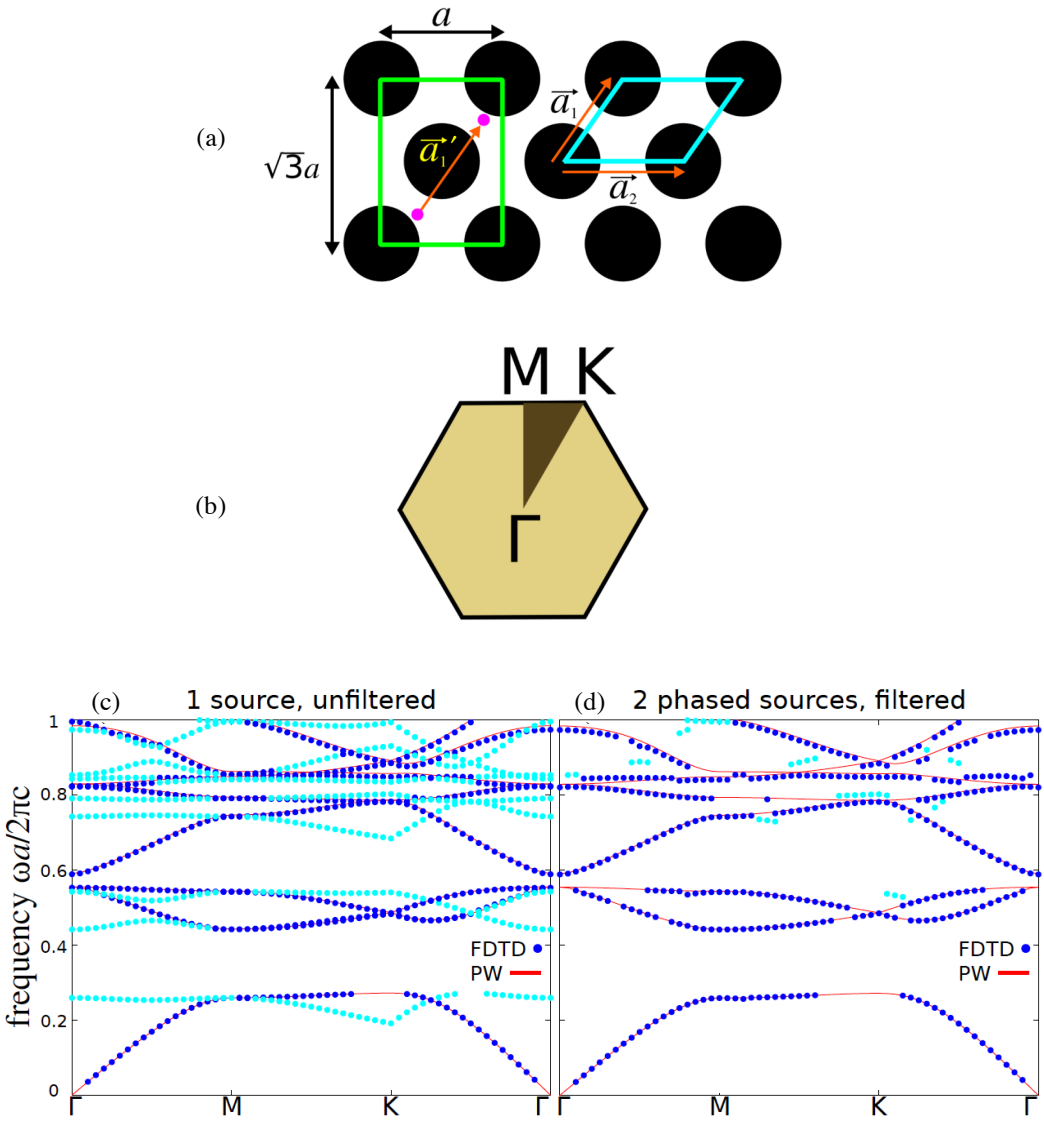

**Fig. 4.9**  Band-folding effects on the dispersion relation, $\omega_n(\mathbf{k})$, computed for a triangular lattice of dielectric rods in air. (a) Geometry of the lattice of dielectric rods, comparing the rectangular supercell of the lattice to the rhombus unit cell. (b) Boundaries of the irreducible Brillouin zone ($\Gamma$–M–K) in $\mathbf{k}$-space, for out-of-plane $\mathbf{E}$ excitation. (c) Band diagram: solid lines—correct curves computed by a plane-wave method [13]; dark dots—correct results of an FDTD calculation using the rectangular supercell of (a), where the eigenfrequencies, $\omega_n(\mathbf{k})$, are computed by filter diagonalization of the response to a pulsed point dipole; light dots—additional spurious "folded" bands arising from the FDTD calculation. (d) Improved FDTD band diagram results using the same rectangular supercell as in (c), but now most of the spurious solutions are eliminated by using two correctly phased sources separated by the lattice vector, $\vec{a}_1'$, in (a) to account for the underlying periodicity, and filtering the results to those with the correct phase relation. A few spurious solutions (light dots) remain due to discretization errors and signal-processing difficulty in dealing with closely spaced bands. These are easily eliminated by hand.

To obtain the FDTD band diagram results shown in Fig. 4.9(c), we excite the modes for the rectangular supercell of Fig. 4.9(a) with a broadband pulsed dipole, and then apply the filter-diagonalization algorithm to the resulting fields to extract the (lossless) resonant frequencies. This procedure yields a combination of correct results for the desired bands (dark dots) that coincide with the plane-wave results (solid lines), as well as spurious "folded" bands (light dots) due to the supercell.

The solution is conceptually simple. If our computational space were a *single* unit cell, our single point source would correspond to a source in *every* unit cell multiplied by the corresponding $e^{i\mathbf{k}\cdot\mathbf{x}}$ phase factor. Since we are trying to model this situation with a supercell of *two* unit cells, we should have *two* sources [shown as light dots in Fig. 4.9(a)] separated by a lattice vector, $\vec{a}_1'$, of the underlying periodicity, and differing by an $e^{i\mathbf{k}\cdot\mathbf{x}}$ phase. In the exact Maxwell's equations, such a two-point source would excite *only* the correct solutions at $\mathbf{k}$, without any possibility of spurious folded bands.

However, the discretization errors inherent in FDTD inevitably cause some spurious folded bands to be excited using this two-point source approach. While these spurious bands are of small amplitude, they still appear upon implementing filter diagonalization. Fortunately, we can filter most of them out of the results, because filter diagonalization provides both the frequency *and* the complex amplitude of the resonances. If we apply filter diagonalization at one point, we can discard modes having a small amplitude. If we apply filter diagonalization at the two source points (or any two points separated by a lattice vector), we can discard any modes whose amplitudes do not have the $e^{i\mathbf{k}\cdot\mathbf{x}}$ phase relationship.

Figure 4.9(d) shows the results of this two-source plus filtering algorithm. Here, we see that most of the spurious solutions (light dots) have been eliminated. There are still a few remnants of the folded bands, because the filter diagonalization technique is not magic — there is still some difficulty in the signal processing in separating the amplitudes of closely spaced modes. In addition, a few real FDTD solutions (dark dots) are also missing. Again, the signal processing is imperfect and occasionally misses resonances, especially if, by bad luck, they are excited with low amplitude.

It is certainly possible to mitigate these problems. For example, we know that the actual bands, $\omega_n(\mathbf{k})$, are continuous, so any gaps must be missing resonances, and any isolated dots must be spurious ones. We could also look more closely at the problematic regions with a narrow-band source, run FDTD for a longer time, or be more clever in the filtering criteria. However, the overall supercell technique described above is generally sufficiently robust that the few remaining errors in the band diagram can be spotted visually and removed manually.

Reiterating, it is more efficient to use the true unit cell via skewed boundary conditions [57, 58]. This also eliminates all spurious modes, but not all signal-processing difficulties for missed resonances or closely spaced modes. However, the use of a supercell with phased sources is a useful work-around that allows an existing FDTD program to be used *with no modifications*.

## 4.7 MOVING SOURCES

An interesting example of a source term is a source whose position is *moving* over time, for example to describe the radiation emitted by a particle as it moves through a medium. A moving *charged* particle can generate *Cherenkov radiation* [59], and a moving dipole antenna (e.g., a moving atom generating spontaneous emission) exhibits Doppler shifts in the frequency of its radiated waves [39].

In a homogeneous medium, the Cherenkov-radiation and Doppler-shift phenomena related to moving sources can be described analytically. However, both phenomena can be greatly modified when the source passes through an inhomogeneous medium. For example, in a homogeneous medium, a moving charge only produces Cherenkov radiation when it exceeds the phase velocity of light in the surrounding medium [59]. In contrast, Cherenkov radiation can be produced at any particle velocity in a periodic medium (the *Smith-Purcell effect* [60, 61]), in addition to a number of other anomalous effects [62, 63]. Similarly, the usual direction of the Doppler shift can be reversed in an inhomogeneous medium, among other unusual effects [64, 65]. FDTD provides a powerful tool to model these effects in complicated media [62, 64].

For example, consider the case of a point charge, $q$, moving with a velocity $\boldsymbol{v}$, which is described by a free-charge density, $\rho = q\delta(\boldsymbol{x} - \boldsymbol{v}t)$. To obtain the equivalent current density, $\boldsymbol{J}$, we turn to the continuity equation (describing conservation of charge) [39]: $\partial \rho / \partial t = -\nabla \cdot \boldsymbol{J}$. Since $\partial \rho / \partial t = -q\boldsymbol{v} \cdot \nabla \delta(\boldsymbol{x} - \boldsymbol{v}t)$, we immediately find that the continuity equation is satisfied by $\boldsymbol{J}(\boldsymbol{x}, t) = q\boldsymbol{v}\delta(\boldsymbol{x} - \boldsymbol{v}t)$. This is a *moving dipole current* oriented in the $q\boldsymbol{v}$ direction.[9,10]

Figure 4.10 illustrates the simulation of such a moving current in FDTD. Here, the current, which is treated as an ordinary $\boldsymbol{J}$ source term in Maxwell's equations, is assumed to move to the right with a superluminal velocity, $v = 1.05 c/n$, where $n$ is the refractive index of the medium. This is equivalent to 0.35 pixel per time-step, $\Delta t$. At every time-step, *we simply change the location of $\boldsymbol{J}$ in the grid* in accordance with the assumed velocity [62]. An immediate complication arises: Because the computation space is discretized in pixels of size $\Delta x$, the successive locations of $\boldsymbol{J}$ would fall exactly on grid-points only if $\Delta x$ is selected to be an integer multiple of $v\Delta t$. This is not the case in this example, and cannot be expected, in general.

One option to resolve this issue is to simply round the location of $\boldsymbol{J}$ to the nearest grid-point at every time-step, as shown in Fig. 4.10(b). However, this results in a jerky, discretized motion of $\boldsymbol{J}$ that generates spurious high-frequency components clearly visible in the radiated field. Instead, in Fig. 4.10(a), we use the interpolation scheme of Chapter 20, Section 20.3.2, to distribute the point-dipole, $\boldsymbol{J}$, to its neighboring grid-points at each time-step, with weights that change continuously with position (and with the correct total $\boldsymbol{J}$). This results in significantly fewer artifacts in the radiated field, although the effects of numerical dispersion are visible in the high-frequency components.

Doppler radiation from a moving dipole source (e.g., to model spontaneous emission from a moving atom) is even more straightforward to implement. A stationary dipole, $\boldsymbol{p}$, oscillating with frequency, $\omega$, is a current, $\boldsymbol{J} = -i\omega\boldsymbol{p}\delta(\boldsymbol{x})e^{-i\omega t}$. The same source moving with velocity, $\boldsymbol{v}$, is $\boldsymbol{J} = -i\omega\boldsymbol{p}\delta(\boldsymbol{x} - \boldsymbol{v}t)e^{-i\omega t}$. [Again, this is nonrelativistic; the exact relativistic formula multiplies the component of $\boldsymbol{J}$ in the direction of $\boldsymbol{v}$ by $(1 - v^2/c^2)^{-1/2}$ [39].] Therefore, as above, one simply has an ordinary dipole source in FDTD whose location changes every time-step [64], using interpolation to achieve continuous "motion." Figure 4.11 illustrates this for the simple case of an oscillating dipole moving at $|\boldsymbol{v}| = 0.3c$ in vacuum. The Doppler shifts (shorter wavelength in front of the source and longer wavelength behind it) are apparent in the FDTD results.

---

[9]Since $\delta(\boldsymbol{x} - \boldsymbol{v}t) = (2\pi)^{-3}\int e^{i\boldsymbol{k}\cdot(\boldsymbol{x}-\boldsymbol{v}t)}d^3\boldsymbol{k}$, upon taking the Fourier transform of this current, one immediately obtains a phase-matching condition, $\omega(\boldsymbol{k}) = \boldsymbol{k} \cdot \boldsymbol{v}$, that must be satisfied by the dispersion relation, $\omega(\boldsymbol{k})$, of any radiated field [59].

[10]Technically, this is a nonrelativistic approximation. The exact relativistic formula multiplies both the rest charge, $q$, and our current, $\boldsymbol{J}$, by $(1 - v^2/c^2)^{-1/2}$ [39].

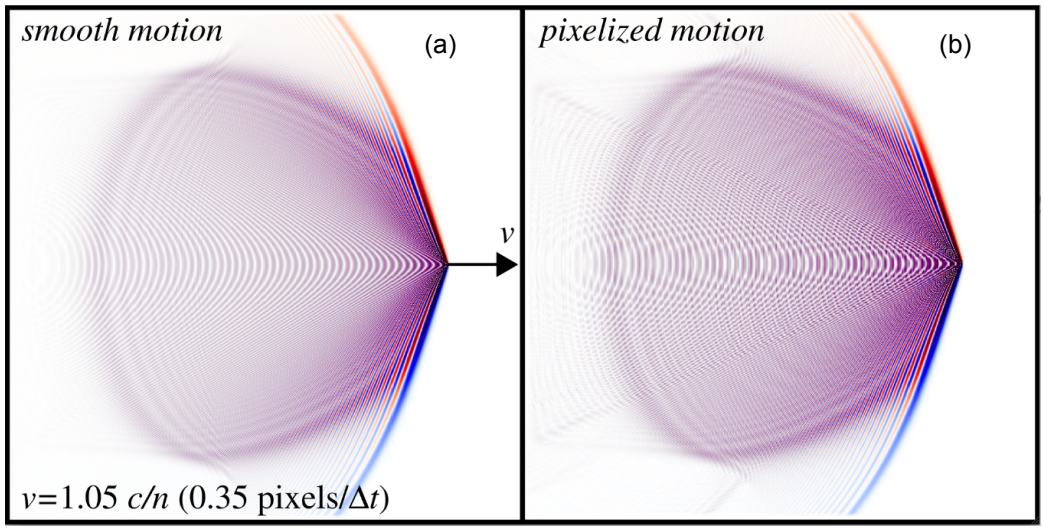

**Fig. 4.10** Visualization of the FDTD-computed Cherenkov radiation ($H_z$ field) from a point charge, $q$, moving to the right with a superluminal velocity, $v = 1.05\,c/n$, in a homogeneous medium, where $n$ is the refractive index of the medium. This is simulated in FDTD by a moving current source, $\boldsymbol{J}(\boldsymbol{x},t) = q\boldsymbol{v}\delta(\boldsymbol{x}-\boldsymbol{v}t)$, where $v$ is equivalent to 0.35 pixel per time-step. (a) Continuous interpolation of the current source onto the FDTD grid, using the technique of Chapter 20, Section 20.3.2. (b) Pixelized motion, in which the location of $\boldsymbol{J}$ is rounded to the nearest FDTD grid-point, leading to visible high-frequency artifacts in the radiated field.

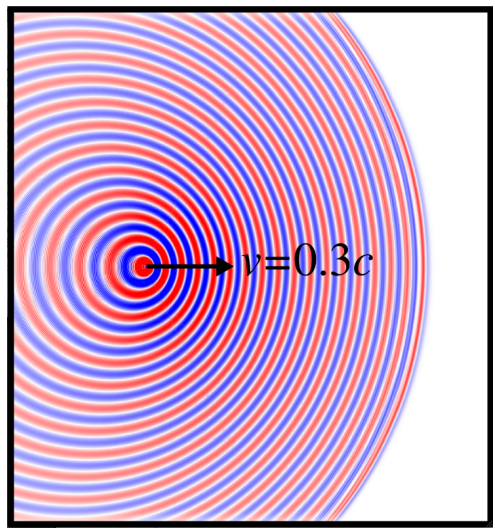

**Fig. 4.11** Visualization of the FDTD-computed Doppler-shifted radiation (out-of-plane electric field) from an oscillating dipole moving to the right at $|\boldsymbol{v}| = 0.3c$ in vacuum, polarized out-of-plane. The waves in front of the source are Doppler-shifted to a higher frequency (shorter wavelength), and the waves behind the source are Doppler-shifted to a lower frequency (longer wavelength).

## 4.8 THERMAL SOURCES

Although sources in FDTD simulations are usually deterministic, there are important cases in which the *physical* source terms in Maxwell's equations are *nondeterministic* currents resulting from thermal and quantum fluctuations. Examples include thermal radiation (radiation from thermal fluctuations in polarizable matter [66–68]), van der Waals and Casimir forces (net forces resulting from imbalances in thermal and quantum radiation pressure [69, 70]), and fluorescence (spontaneous emission from excited particles).

Because one is generally interested only in the time-average result of these fluctuations, there are usually clever ways to employ FDTD (or another Maxwell's equations solver) to obtain the averaged result without modeling the fluctuations directly. For example, far-field thermal radiation obeys *Kirchhoff's law*, which says that the thermal radiation from a body is equal to the known radiation of an ideal "black body" multiplied by the absorptivity (fraction of absorbed power) of the body at each frequency [68]. In fact, the fraction of absorbed power from an incident plane wave is easily computed in FDTD (e.g., by the techniques of Sec. 4.5). More generally (e.g., in the near field), one can employ techniques derived from electromagnetic reciprocity and related principles [71–74]. For example, the rate and extraction efficiency of spontaneous emission can be computed by the power radiated from a dipole [24–26], as described in Section 4.4. Furthermore, as discussed in detail in Chapter 19, time-average Casimir forces can be computed efficiently in FDTD using reciprocity and other techniques [75, 76]. It is also possible to efficiently model time-average blackbody radiation and electromagnetic fluctuations in dissipative open systems using FDTD, as discussed in detail in Chapter 18.

Nevertheless, directly modeling fluctuating currents in FDTD — a type of *Monte Carlo method* or *Langevin model* — has the virtues of simplicity and generality. This approach is attractive when studying new problems where more sophisticated methods are not yet implemented. For example, the use of fluctuating sources in FDTD was the first method successfully employed in studying *near-field* thermal radiation (radiative heat transport between bodies at such small separations that evanescent interactions become important [77]) for any geometry other than spheres or planes [71]. Moreover, the elementary nature of the fluctuating-currents picture may make it more approachable than more sophisticated formulations.

Consider a material with a complex permittivity tensor, $\varepsilon(\omega)$, at a temperature, $T$. Physically, $\varepsilon \neq \varepsilon_0$ corresponds to a *polarizable* medium, in which microscopic dipoles can be aligned (or created) in response to an applied electric field [39]. However, even in the absence of an applied electric field, thermal and quantum fluctuations induce *spontaneous* microscopic polarizations in the material, which rapidly fluctuate in orientation and magnitude. This fluctuating polarization density, $P$, corresponds to a fluctuating *current* density, $J = \partial P / \partial t$ [39], with zero mean, $\langle J \rangle$, and nonzero mean-square, $\langle |J|^2 \rangle$. As a result, fluctuating electromagnetic fields are generated, as depicted in Fig. 4.12(b).

There is an important consequence if we Fourier transform these current-density fluctuations to obtain $\hat{J}(\omega)$. Namely, the mean-squared current, $\langle |\hat{J}|^2 \rangle$, is proportional to the absorption coefficient, Im $\varepsilon$, of the material. This is a result of a profound and far-reaching principle of statistical physics called the *fluctuation–dissipation theorem* [67]. In particular, the statistics of the fluctuation in $J$ are described by [67]:

$$\langle \hat{J}_\ell(\omega, x)\, \hat{J}_m(\omega, x')^* \rangle = \frac{1}{\pi} \delta(x - x') \left[ \frac{\hbar\omega}{2} \coth\left( \frac{\hbar\omega}{2kT} \right) \right] \omega\, \text{Im}\, \varepsilon_{\ell m}(\omega, x) \qquad (4.31)$$

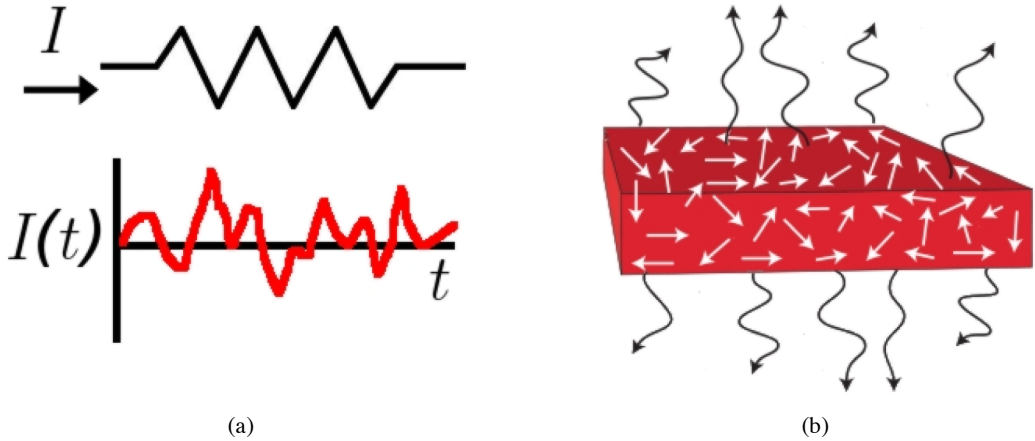

**Fig. 4.12** Schematic effects of random thermal and quantum fluctuations in polarizable matter. (a) At a macroscopic level, these fluctuations produce Johnson–Nyquist noise: random current fluctuations through a resistor. (b) At a microscopic level, the fluctuations can be described semiclassically as a fluctuating current density, $\boldsymbol{J}$, whose statistics are given by (4.31) from the fluctuation-dissipation theorem. This current produces thermal radiation, van der Waals forces, and other phenomena.

where $\hbar$ is Planck's constant and $k$ is Boltzmann's constant. In signal-processing language, the right-hand side of (4.31) is the power spectral density of a "colored noise" process, i.e., the Fourier transform of the correlation function of the currents in time.

It is instructive to examine the right-hand side of (4.31) term by term. The first term, $\delta(\boldsymbol{x}-\boldsymbol{x}')$, means that the fluctuations are *uncorrelated in space*.[11] The second term (in $[\cdots]$) is the spectrum of the thermal and quantum fluctuations. For low frequencies or high temperatures ($\hbar\omega \ll 2kT$), this spectrum becomes simply $kT$, independent of frequency, since $\coth x \approx 1/x$ for small $x$. In fact, this is the *classical* limit. Alternatively, at high frequencies or low temperatures, $\coth \approx 1$, and the spectrum approaches $\hbar\omega/2$. This is the interesting regime of the purely quantum-mechanical phenomenon of *zero-point fluctuations* [67]. Finally, the third term, $\omega \operatorname{Im}\varepsilon$, corresponds to the *conductivity* of a conducting material [39].

This analysis leads us to a famous result in electrical engineering, the *Johnson–Nyquist* noise formula [79]. If we first integrate the current density over the cross-section of a wire to obtain the mean-square current, $\langle \tilde{I}^2 \rangle$, then integrate over a frequency bandwidth, $\Delta\omega = 2\pi\Delta f$ (summing over both positive and negative frequencies), and finally consider the classical limit, $[\cdots] \approx kT$, we arrive at the famous formula, $\langle I^2 \rangle = 4kT\Delta f/R$, where $R$ is the wire's resistance. Thus, (4.31) is simply the microscopic generalization, including quantum effects, of Johnson–Nyquist noise, which produces a fluctuating current in any resistor as depicted in Fig. 4.12(a).

---

[11]Technically, the fluctuations are uncorrelated in space for the usual case of a material with a *local* dielectric response. Materials having a *nonlocal* dielectric response, in which a field at one point produces a polarization at a different point, produce correlated fluctuations. This property can be modeled in FDTD [78], as described in detail in Chapter 9.

To implement such a fluctuating current source in FDTD, one could in principle employ sources described exactly by (4.31). However, there is a useful technique that greatly simplifies the algorithm and improves its performance [71, 80, 81]. Physically, (4.31) simply describes a simulation in which there is an independent random current source at every point in space, or at least, every point where $\text{Im}\,\varepsilon > 0$. From the resulting fields, we can compute the time-average energy flux (Poynting vector), momentum flux (stress tensor), or any other desired quantities. The biggest computational difficulty is that, while the sources are uncorrelated in space, they are correlated in time with power-spectrum given by (4.31). There are various ways to generate correlated random numbers (colored noise) by filtering techniques from signal processing, but they involve substantial additional storage or computation, or both [82].

Instead, we can exploit the *linearity* of electromagnetism, assuming linear materials. Namely, we can instead compute the fields due to *white-noise* currents (uncorrelated in time and space), and then only *after* the simulation, multiply the Fourier-transformed fields by the $\omega$-dependent terms of (4.31). We begin by considering the common case in which there is only *one* absorbing material, so that $\text{Im}\,\varepsilon$ is either zero (e.g., in air regions) or some function of frequency, $\varepsilon''(\omega)$, in the absorbing material. Then, we inject white noise currents in the absorbing material. Here, each component of $\boldsymbol{J}$ in the absorber is an uncorrelated random number generated at each FDTD time-step, with mean, 0, and mean-square, 1. Thanks to the central-limit theorem, the precise random-number distribution is irrelevant, e.g., Gaussian or uniform. As long as the distribution has the correct mean and mean-square, the net effect is the same when averaged over a long time, or over many simulations.

One then computes the desired Fourier-transformed fields (e.g., for the Poynting flux). Because one usually wants only the average of these quantities, one should compute the "ensemble average" over many such simulations, or over a long time. (Technically, such an average is called a periodogram; there are various windowing algorithms to speed convergence [17]). *After* this is done, the Poynting spectrum (or any quantity proportional to the squared Fourier amplitudes) is multiplied by the $[(\hbar\omega/2)\coth(\hbar\omega/2kT)]\omega\varepsilon''(\omega)$ frequency dependence to obtain the correct "thermal" spectrum.

Figure 4.13 illustrates the type of result that may be obtained using this method, excerpted from more detailed results described elsewhere [71]. The solid curve in the figure plots the spectral density of the near-field power flux between two one-dimensionally periodic photonic crystals of silicon carbide (SiC) separated by a short distance, $d$ (inset). The dashed curve plots the power flux between *unpatterned* SiC slabs. In both cases, the power flux is normalized by the flux between the same structures at infinite separation, $d \to \infty$. The patterning of the slabs drastically modifies the flux spectrum as compared to the unpatterned case. In this case, however, there is a complication. Namely, we have two bodies at *different* temperatures, $T$. This would seem to conflict with our technique of multiplying by the $\coth(\hbar\omega/2kT)$ temperature dependence only after the simulation.

There are two possible solutions. Because the currents are uncorrelated in space, we could simulate them *separately* (the cross-terms in the resulting fields average to zero). This would mean two simulations, one with sources in each body, with the results of each simulation multiplied by its respective coth factor and summed. However, as a consequence of electromagnetic reciprocity [71], the energy flux into body 1 in response to white-noise sources in body 2 is *identical* to the energy flux into body 2 from white-noise sources in body 1. Therefore, it suffices to perform *one* simulation, with white-noise sources in only *one* body, and then multiply the results by the *difference* of the two coth factors to obtain the net energy flux [71].

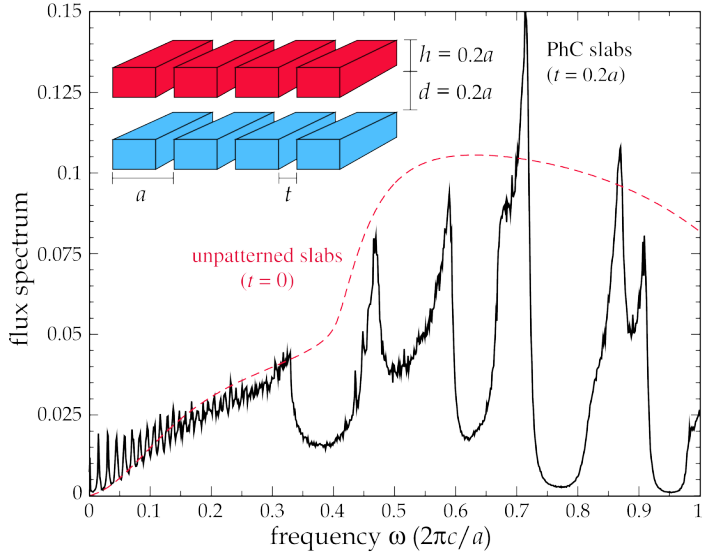

**Fig. 4.13** FDTD-computed radiative heat transfer between patterned and unpatterned SiC slabs, operating in the *near field* where the slabs are close enough to couple evanescently as well as radiatively [71]. The solid curve plots the spectral density of the power flux between SiC photonic crystals (inset) maintained at unequal temperatures and surface-to-surface separation, *d*. The dashed curve plots the power flux between unpatterned SiC slabs. In both cases, the power flux is normalized by the power flux that would exist between the same structures at infinite separation, $d \to \infty$.

There is one additional technique that is required if there are *multiple* absorbing materials in a single simulation, each with its own absorption spectrum, Im $\varepsilon$. In this case, we cannot use white-noise sources in all absorbing materials at once and multiply by the frequency-dependence *a posteriori*, since the fields from different absorbing materials are mixed together.

A potential approach to this problem would be to perform a separate simulation for each material (using white-noise sources in one material at a time), and summing them afterwards. This would work because of the lack of spatial correlation. However, there is another possibility that would use just a *single* simulation, based on the way that a frequency-dependent $\varepsilon$ is modeled in FDTD. In particular, a frequency-dependent permittivity, $\varepsilon(\omega) = \varepsilon_0[1 + \chi(\omega)]$, is typically implemented in FDTD by integrating an *auxiliary differential equation* [1]. By inserting the white-noise source into this auxiliary equation *instead* of directly into $J$, one can build the desired Im $\varepsilon$ frequency dependence directly into the current, up to material-independent factors of $\omega$ that can be multiplied *a posteriori* [80, 81].

## 4.9 SUMMARY

This chapter has provided a tutorial discussion of the state of knowledge regarding the relationships between current sources and the resulting electromagnetic waves in FDTD simulations. The techniques presented here are suitable for a use in a wide range of FDTD modeling applications, from classical radiation, scattering, and waveguiding problems to nanoscale material structures interacting with thermal and quantum fluctuations.

The chapter commenced with a discussion of incident fields and equivalent currents, examining in detail the principle of equivalence and the discretization and dispersion of equivalent currents in FDTD models. This was followed by a review of means to separate incident and scattered fields, whether in the context of scatterers, waveguides, or periodic structures. The next major topic was the relationship between current sources and the resulting local density of states. Here, key sub-topics included the Maxwell eigenproblem and the density of states, radiated power and the harmonic modes, radiated power and the local density of states, computation of the local density of states in FDTD, Van Hove singularities in the local density of states, and resonant cavities and Purcell enhancement. Subsequent major topics included source techniques that enable covering a wide range of frequencies and incident angles in a small number of simulations for waves incident on a periodic surface; sources to efficiently excite eigenmodes in rectangular supercells of periodic systems; moving sources to enable modeling of Cherenkov radiation and Doppler-shifted radiation; and finally thermal sources via a Monte Carlo/Langevin approach to enable modeling radiative heat transfer between complex-shaped material objects in the near field.